\documentclass[twocolumn,secnumarabic,amsmath,amssymb,showpacs,nobibnotes,aps, reprint,prl]{revtex4-1}

\usepackage{amssymb}

\usepackage{graphicx}
\usepackage{float}
\usepackage{dcolumn}
\usepackage{bm}
\usepackage{color}
\usepackage{caption}

\begin{document}

\title{Evolution of self-organized structure, the internal transport barrier in the ion-temperature-gradient driven gyrokinetic turbulence}

\author
{Shaojie Wang$^{\ast\dagger}$, Zihao Wang$^{\dagger}$, Tiannan Wu$^{\dagger}$\\
{$^\ast$ Corresponding author. Email: wangsj@ustc.edu.cn.}
{$^\dagger$ These authors contributed equally to this work.}
}


\affiliation{Department of Engineering and Applied Physics, University of Science and Technology of China, Hefei, 230026, China}
\date{\today}

\begin{abstract}
Understanding the self-organization of the most promising internal transport barrier  in fusion plasmas needs a long-time nonlinear gyrokinetic global simulation. The Neighboring Equilibrium Update method is proposed, which solves the secularity problem in a perturbative simulation and speeds up the numerical computation by more than 10 times. It is found that the internal transport barrier emerges at the magnetic axis due to inward propagated turbulence avalanche, and its outward expansion is the catastrophe of self-organized structure induced by outward propagated avalanche.   

\end{abstract}

\maketitle

Self-organized states\cite{BakPRL87,BakPRA88} have been found in many nonlinear complex systems, such as the rainforest in ecology\cite{SoleJTB95}, the earthquakes in geophysics\cite{SornetteJGR90}, the laser interaction with matters in optical engineering\cite{NayakOLE10}, the nanotubes in electroanalytical chemistry\cite{MacakJEC08}, and the turbulence in magnetic fusion plasmas\cite{KishimotoPoP96}. Following Bak-Tang-Wiesenfeld\cite{BakPRA88}, by ``self-organized'', one means that the system naturally evolves to the state, insensitive to the initial conditions.
The International Thermonuclear Experimental Reactor (ITER)\cite{ShimadaNF07}, a tokamak fusion torus, will be a milestone in magnetic fusion energy research\cite{OngenaNP16,FasoliPRL23}, whose success crucially depends on the core plasma confinement improvement, designated by the formation of Internal Transport Barrier (ITB)\cite{DoyleNF07}. 
Various ITBs have been found in tokamaks, such as JT-60U\cite{Koide-JT60U-PRL94,Fujita-JT60U-PRL01}, TFTR\cite{Levinton-TFTR-PRL95,SynakowskiPRL97}, DIII-D\cite{Strait-DIIID-PRL95,BurrellPPCF98,RettigPoP98}, JET\cite{Crisanti-JET-PRL02,MazziNP22}, ASDEX-U\cite{Hobirk-ASDEX-PRL01}, HL-2A\cite{Yu-HL2A-NF16}, EAST\cite{LiPRL22,YeNF22} and KSTAR\cite{HanNature22} due to different turbulence-reduction effects\cite{ConnorNF04,Mantica-JET-08,IdaPPCF18}, such as the radial electric field ($E_r$) shearing\cite{BiglaryPFB90}, the negative or weak magnetic shear\cite{RomanelliPoP93,KishimotoPPCF98,RogisterPoP00,ConnorPPCF04,Levinton-TFTR-PRL95}, the external momentum injection\cite{SynakowskiPRL97,Sakamoto-JT60U-NF01}, and the effects of energetic ions\cite{HanNature22,MazziNP22}. The most promising ITB for ITER\cite{DoyleNF07} is the one emerging near the magnetic axis and expanding radially outward in a weak/positive magnetic shear heated plasma without momentum injection\cite{Koide-JT60U-PRL94,BurrellPPCF98,RettigPoP98}, because the magnetic configuration in this hybrid scenario\cite{DoyleNF07,FasoliPRL23} is relatively easier to control, and the formation of the ITB seems to be nonlinearly self-organized\cite{ConnorNF04,IdaPPCF18}. Although the most promising ITB has been observed in many tokamaks\cite{Koide-JT60U-PRL94,BurrellPPCF98,Yu-HL2A-NF16,LiPRL22,HanNature22,MazziNP22}, its formation dynamics has not been well understood. 

Due to the complexity and nonlinearity, the nonlinear gyrokinetic (GK) simulation\cite{LinScience98,HanNature22,MazziNP22} has become indispensable in turbulent transport research, which is critical in understanding the ITB physics. Local simulations\cite{HanNature22,MazziNP22} are not sufficient for investigating the non-local effects, such as the turbulence avalanche\cite{McMillanPoP09,StrugarekPRL13} and the ITB expansion. 
Therefore, it is of significant interest to make a nonlinear GK global simulation to investigate the formation dynamics of the most promising ITB. 
To solve this challenging problem, we need a nonlinear GK simulation including the magnetic axis where the ITB emerges. 
Many efforts have been made on the nonlinear GK global simulation, which has led to the discovery of Zonal Flows (ZFs)\cite{LinScience98,ChenPoP00,DiamondPRL94,WangPoP17} nonlinearly excited by the ion-temperature-gradient (ITG) mode and reducing the turbulence. The nonlinear GK global codes NLT\cite{DaiCPC19}, GT5D\cite{Idomura-GT5D-CPC08}, ORB5\cite{Jolliet-ORB5-CPC07} and GKNET\cite{ImaderaPPCF23}, have been developed to include the magnetic axis. However, it is still difficult to simulate a realistic formation process of the ITB. In a long-time simulation, the $\delta f$ codes may involve the secularity problem; the Eulerian codes may becomes too slow due to the CFL constraint, since the ZF may become strong. The first nonlinear GK global simulation of ITB formation was carried out by using the semi-Lagrangian code GYSELA\cite{StrugarekPRL13}, which did not include the magnetic axis, and needed an external injection of vorticity. 
More recently, a nonlinear GK global simulation by the full-$f$ Eulerian code, found an ion-ITB formed with a very localized external momentum injection in a hybrid scenario configuraion\cite{ImaderaPPCF23}. The computation cost in this GK simulation is extremely high; the time step used in a usual GK simulation is larger than the period of ion gyro-motion, $\tau_{gy}$, however, in Ref. \onlinecite{ImaderaPPCF23}, it is reduced to $\sim 0.13\tau_{gy}$ due to the CFL constraint; therefore the simulation domain is reduced to a quarter-torous there. More importantly, these simulated ITBs\cite{StrugarekPRL13,ImaderaPPCF23} are not the most promising ITB, since they critically depend on the $E_r$ shearing externally driven by either the vorticity or momentum injection. 
Therefore, it is of significant interest to further develop the method of long-time nonlinear GK global simulation to investigate the formation dynamics of the most promising ITB (hereafter, it will be simply noted as the ITB).

Here we report the Neighboring Equilibrium Update (NEU) method, which solves the secularity problem for a $\delta f$ code, and significantly speeds up the present long-time nonlinear GK simulation. With the NEU method, we have successfully carried out for the first time a nonlinear GK global simulation of the formation of the ITB, which reveals that the expansion of the ITB is a catastrophe of the self-organized structure induced by turbulence avalanche. 

We simulate the ITG turbulence with adiabatic electrons and kinetic ions satisfying the nonlinear GK equation\cite{BrizardRMP07},
\begin{equation}\label{eq:GKE-full-f}
\partial_t f +\{H,f\}=\mathcal{S}+\mathcal{C}(f),
\end{equation}
with $f(\bm{z},t)=f\left(r,\theta,v_{\|},\mu,\alpha,t\right)$ the distribution function of ion gyro-centers, $\{,\}$ the Poisson bracket, $\mathcal{S}$ the ion heating term, $\mathcal{C}(f)$ the ion-ion collision term. The neoclassical ion thermal conductivity given by the simulation here is typically $\sim0.2\mathrm{m^2/s}$, in good agreement with the theory\cite{WessonBook97}. Here $r$ and $\theta$ are minor radius and poloidal angle of the torus, respectively; $r=r(\psi)$, with $\psi$ the poloidal magnetic flux. $\alpha=q\theta-\zeta$, with $q$ the safety factor and $\zeta$ the toroidal angle. $v_{\|}$ and $\mu$ are parallel velocity and magnetic moment, respectively. 
The gyro-center Hamiltonian is  
$H=H_0+\delta H$, with the equilibrium $H_0=m_iv_{\|}^2/2+\mu B +e_i\langle \Phi_0 \rangle_{g}$ and the perturbation $\delta H=e_i\langle \delta\Phi \rangle_{g}$. 
Here $e_i$ and $m_i$ are the ion charge and mass, respectively; $B$ is the magnetic field. $\Phi_0$ and $\delta\Phi$ are equilibrium and perturbed electrostatic potential, respectively; $\langle \cdot \rangle_{g}$ is the gyro-average operator. Eq. (\ref{eq:GKE-full-f}) is used in the full-$f$ method. 
In the $\delta f$ method, the full distribution function is separated into $f=f_0+\delta f$, 
with the equilibrium distribution, $f_0$, defined by
\begin{equation}\label{eq:f_0-standard}
\{H_0,f_0\}=0.
\end{equation}
Subtracting it from Eq. (\ref{eq:GKE-full-f}) yields the GK $\delta f$ equation
\begin{equation}\label{eq:GKE-delta-f}
\partial_t \delta f +\{H_0,\delta f\}=-\{\delta H,f_0\}-\{\delta H, \delta f\}+\mathcal{S}+\mathcal{C}. 
\end{equation}
In the usual $\delta f$ method which solves Eq. (\ref{eq:GKE-delta-f}), the time-independent $f_0$ is taken approximately as a local Maxwellian, which is not exactly a constant of motion\cite{WangPRE01,IdomuraNF03}. 

The Non-Linear Trubulence (NLT) code, which has been benchmarked with various codes\cite{YeJCP16,XuPoP17,DaiCPC19}, is used here to solve Eq. (\ref{eq:GKE-delta-f}); it evolves $\delta f$ along the equilibrium orbit by using the characteristic line method and takes account of the perturbation effects by using the Numerical Lie-Transform\cite{WangPoP12,WangPoP13}. To solve the GK quasi-neutrality equation for the ITG fluctuations, we use the 8-point gyro-average method \cite{LeeJCP87}, while for ZFs, we still use the long-wave-length approximation\cite{BrizardRMP07}. 

The $\delta f$ method has a higher numerical precision. However, in a long-time simulation, successive nonlinear neighboring equilibrium\cite{ChenNF07,BottinoJoP22} is formed; $\delta f$ may become too large to keep the high precision. To avoid this problem, we propose the NEU method, by updating the equilibrium ($H_0$ and $f_0$) of the system, namely, changing the partitions from  $H=H_0+\delta H$, $f=f_0+\delta f$ to
\begin{equation}\label{eq:NEU-update}
 H=\overline{H_0}+\overline{\delta H},~
 f=\overline{f_0}+\overline{\delta f},
\end{equation}
to keep $\overline{\delta H}$ and $\overline{\delta f}$ small. 

Define the ensemble average of a scalar function $g(\bm{z},t)$ as $g_{en}(\bm{Z})=\frac{1}{\tau_{en}}\int_{\tau_0-\tau_{en}}^{\tau_0}\mathrm{d}t\frac{1}{2\pi}\oint \mathrm{d}\alpha f(\bm{z},t)$, with $(\bm{Z})=(r,\theta,v_{\|},\mu)$, and $\tau_{en}\approx 10R/c_s$; here $m_ic_s^2=T_{e,0}$, with $T_{e,0}$ the central electron temperature. The updated equilibrium is given by $\overline{H_0}$ and $\overline{f_0}$. $\overline{H_0}(\bm{Z})= H_{en}(\bm{Z})$.
\begin{equation}\label{eq:f_0-construction}
\overline{f_0}(\bm{Z})=\tau_b^{-1}\oint \mathrm{d}\tau  f_{en}\left[\mathcal{\bm{Z}}(\tau;\bm{Z},\tau_0)\right],
\end{equation}
with $\tau_b=\oint \mathrm{d}\tau$; the integral is taken over the poloidally closed orbit determined by $\overline{H_0}$; $\mathcal{\bm{Z}}(\tau;\bm{Z},\tau_0)$ is the phase space point at $t=\tau$ on the orbit of the gyro-center launched at $t=\tau_0$ from $\bm{Z}$, $\mathcal{\bm{Z}}(\tau_0;\bm{Z},\tau_0)=\bm{Z}$.  
By definition, $\overline{f_0}\left[\mathcal{\bm{Z}}(\tau;\bm{Z},\tau_0)\right]=\overline{f_0}(\bm{Z})$, $\overline{f_0}$ is a constant of motion along the orbit. Since $\{\overline{H_0},\overline{f_0}\}=\dot{\bm{Z}}\cdot\partial_{\bm{Z}}\overline{f_0}$ represents the variation of $\overline{f_0}$ along the orbit given by $\overline{H_0}$, one finds 
\begin{equation}\label{eq:f_0-update}
\{\overline{H_0},\overline{f_0}\}=0.   
\end{equation}
Subtracting Eq. (\ref{eq:f_0-update}) from Eq. (\ref{eq:GKE-full-f}), one finds 
\begin{equation}\label{eq:GKE-delta-f-NEU}
\partial_t \overline{\delta f} +\{\overline{H_0},\overline{\delta f}\}=-\{\overline{\delta H},\overline{f_0}\}-\{\overline{\delta H}, \overline{\delta f}\}+\mathcal{S}+\mathcal{C},
\end{equation}
the updated $\overline{\delta f}$ equation to be solved to advance the system after NEU. Eq. (\ref{eq:GKE-delta-f-NEU}) is formally same as Eq. (\ref{eq:GKE-delta-f}). 

Using the constants of motion, $P_{\alpha}$ (canonical toroidal angular momentum) and $W$ (energy), to change the variables from $(\bm{Z})$ to $(P_{\alpha},W,\mu,\theta)$, one finds $f_{en}(\bm{Z})=F_{en}(P_{\alpha},W,\mu,\theta)$. Using $\mathrm{d}\tau=\mathrm{d}\theta/\dot{\theta}$, one finds Eq. (\ref{eq:f_0-construction}) is reduced to $\overline{F_0}(P_{\alpha},W,\mu)=\tau_b^{-1}\oint \mathrm{d}\theta/\dot{\theta} F_{en}(P_{\alpha},W,\mu,\theta)$, the definition used to diagnose the evolution of $\overline{F_0}$ in ORB5\cite{BottinoJoP22}, where the $\delta f$ equation was not updated. 

In a long-time simulation, we perform the NEU in NLT, whenever the ion temperature is changed by $15\%$ or the CFL constraint, $\delta\bm{g}\ge 0.8\bm{\lambda}_{min}/3$, is touched; here $\delta\bm{g}$ is the perturbed displacement computed by the numerical Lie-transform\cite{WangPoP13} and $\bm{\lambda}_{min}$ is the minimum wave-length in the system. By moving the symmetric radial electric field to the equilibrium, the NEU method significantly relaxes the CFL constraint in the NLT code. 

\begin{figure}[htbp]
\includegraphics[width=8.0cm]{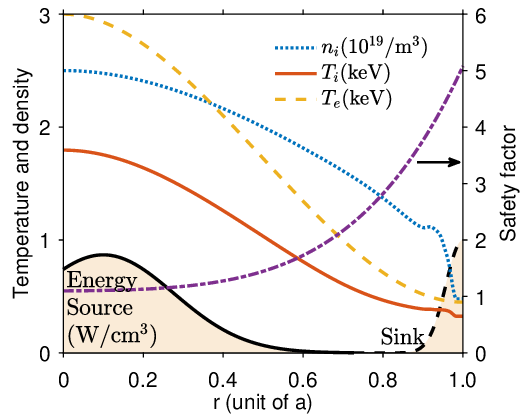}
\caption{~Equilibrium and heating profiles.
}
\label{fig1}
\end{figure}

The main parameters here are chosen to model a DIII-D-like deuterium plasma\cite{BurrellPPCF98}. The major/minor radius of the torus are $R/a=1.67\mathrm{m}/0.67\mathrm{m}$; the toroidal magnetic field is $B_T=2.1\mathrm{T}$. Initial  profiles of ion density $n_i$, ion/electron temperature $T_i$/$T_e$, safety factor $q$, and heating power density, are shown in Fig. 1. The ion heating power is $P=2.5\mathrm{MW}$. A heat sink term is added near the edge ($r>0.9a$ is the buffer region). 
The simulation domain is $r/a\in[0,1]$, $\theta\in[-\pi,\pi]$, $\alpha\in[0,2\pi]$, $v_{\|}/c_s\in[-6,6]$, $\mu B_0/T_{0,e}\in[0,6^2/\sqrt{2}]$; here $B_0=B(r=0)$.
Grid numbers are $N_r\times N_{\theta}\times N_{\alpha}\times N_{v_{\|}}\times N_{\mu}=222\times16\times190\times96\times16$. $\mu$ is discretized according to the Gauss-Legendre formula, while the other variables are discretized uniformly. The time step here is $\Delta t=4\tau_{gy}$, which is 30 times larger than used in Ref. \onlinecite{ImaderaPPCF23}; the NEU method significantly speeds up the computation here. Note that we simulate the entire torus here rather than a quarter-torus\cite{GarbetPoP01,ImaderaPPCF23}. The convergence study has been carried out for this work; for example, when changing $\Delta t$ from $4\tau_{gy}$ to $2\tau_{gy}$, the detail of the results, such as the timing of the burst events in the nonlinear phase, changes indeed, however, the conclusions made in this work do not change qualitatively. 

\begin{figure}[htbp]
\includegraphics[width=8.0cm]{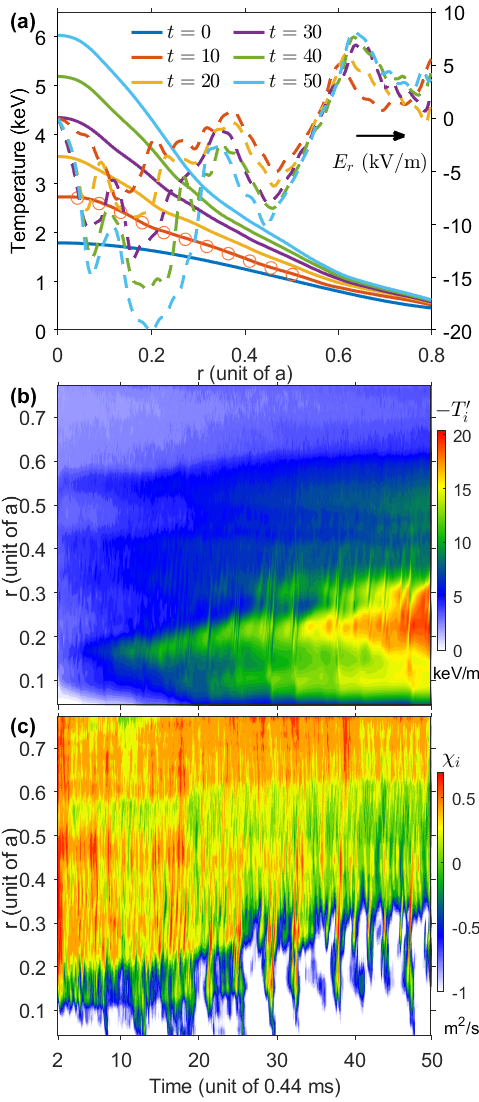}
\caption{~Formation of the ITB in ITG turbulence. 
{\bf(a)} Profiles of $T_i$ and $E_r$ at different time. Open circles: $T_i$ at $t=10$ found by a simulation without NEU. 
{\bf(b)} The temperature gradient, $-T'_i(r,t)$. The ITB emerges around $r\approx 0.16a$ at $t \le  10\times 0.44 \text{ms}$, and its center expands to $r\approx 0.24a$ at $t\approx 50\times 0.44 \text{ms}$.
{\bf(c)} The turbulent thermal conductivity $\chi_i(r,t)$, which also signifies the turbulence intensity. 
}
\label{fig2}
\end{figure}

The general results are shown in Fig. 2. The time is normalized here by $100R/c_s\approx 0.44\mathrm{ms}$. Fig. 2(a) shows that $T_i$ at $t=10\times 0.44\mathrm{ms}$ agrees well with a standard $\delta f$ simulation without NEU; this verifies the NEU method. The following consistencies with previous experimental and theoretical results are demonstrated. (1) The ITB spontaneously emerges near the magnetic axis and radially outward expands in a speed of $\sim 3\mathrm{m/s}$ in a heated plasma with a weak/positive magnetic shear\cite{Koide-JT60U-PRL94,BurrellPPCF98}[Figs. 2(a-c)]. (2) The $E_r$ shear appears at the ITB location\cite{ConnorNF04,IdaPPCF18,BiglaryPFB90}[Fig. 2(a)]. (3) The intermittent burst events on both sides of the ITB\cite{Mantica-JET-08} are observed [Fig. 2(c)].
(4) The successive collapse\cite{StrugarekPRL13} and expansion\cite{Koide-JT60U-PRL94,BurrellPPCF98} of the ITB are observed [Fig. 2(b, c)].
(5) The power threshold behavior\cite{Koide-JT60U-PRL94,BurrellPPCF98,ConnorNF04,IdaPPCF18} is suggested by a simulation with a lower power (0.6 $\mathrm{MW}$) which shows no ITB expansion; this is consistent with the previous simulation\cite{ImaderaPPCF23}, which shows no ITB formed without external momentum injection, with a heating power $4\mathrm{MW}$ and a central particle density $2.5\times10^{20}/\mathrm{m}^3$; the power threshold is proportional to the particle density\cite{ConnorNF04}. These observations validate the simulation here. 

\begin{figure*}[htbp]
\includegraphics[width=17.0cm]{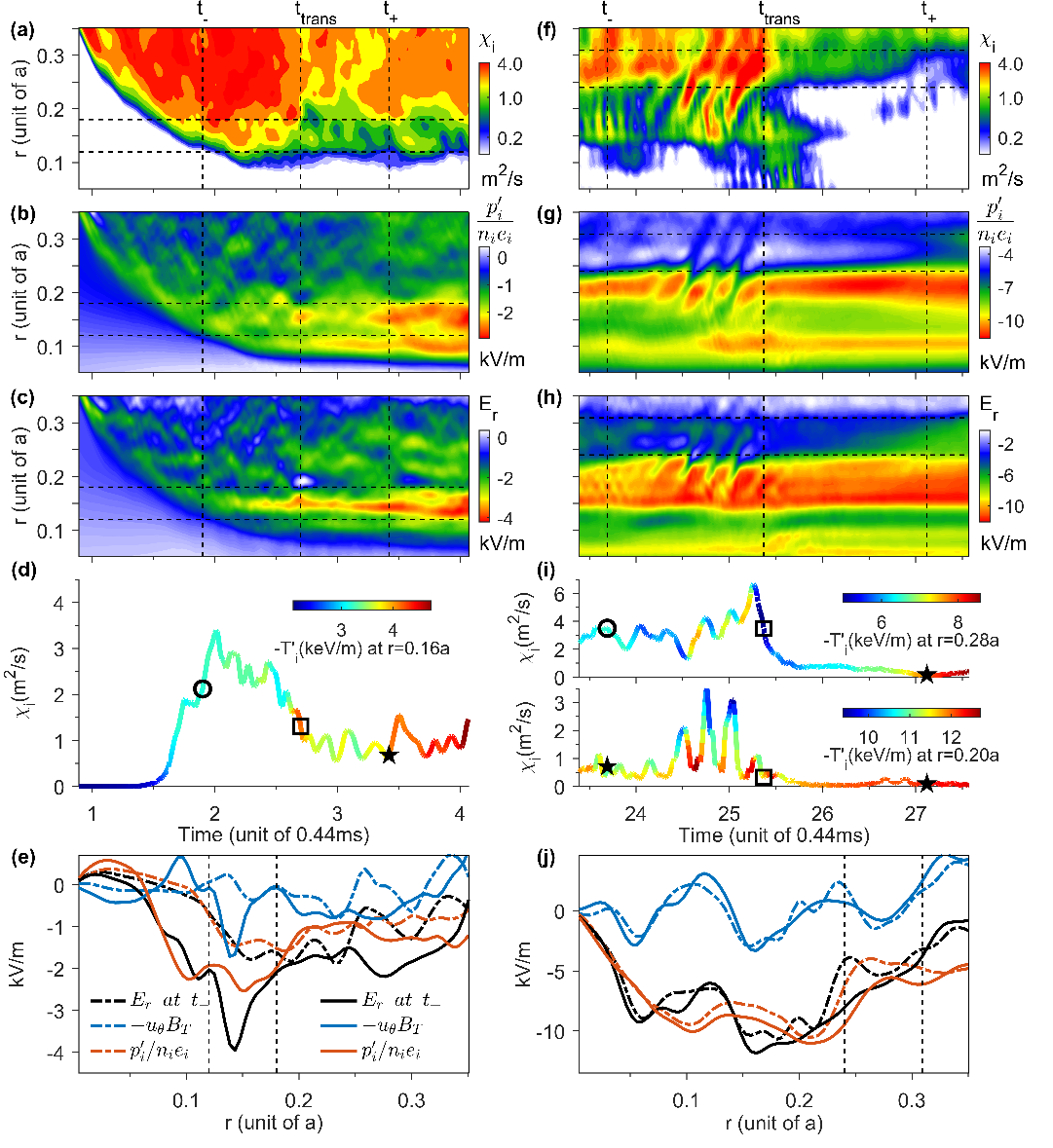}
\caption{~Dynamics of the ITB formation. 
{\bf(a-e)} Emergence at $r=0.16a$. 
{\bf(f-j)} Expansion at $r=0.28a$. $t_{trans}$: transition time (different for the two columns). 
{\bf(a, f)} $\chi_i$. The turbulence is reduced within the mesoscale region marked by the two horizontal lines.
{\bf(b, g)} Ion pressure gradient ($p_i'$).
{\bf(c, h)} $E_r$ profile. 
{\bf(d, i)} The $\chi_i$-gradient relation; open circles: the L-state; the solid stars: the H-state; open squares: transition time. 
{\bf(e, j)} $E_r$ and its contributions before ($t=t_{-}$) and after ($t_{+}$) transition; the effect of the toroidal rotation is negligible here. 
} \label{fig3}
\end{figure*}

The dynamics of ITB formation is shown in Fig. 3. Figs. 3(a-c) indicate that the ITB emergence at $r=0.16a$ is induced by the radially inward propagated avalanche; the direction of this propagation is consistent with the fact that it is more stable near the magnetic axis where the magnetic shear is weaker\cite{RomanelliPoP93,ConnorPPCF04}. The S-curve shown in Fig. 3(d) suggests a transition from a Low-confinement (L) state (high $\chi_i$) to a High-confinement (H) state (low $\chi_i$) when the ITB emerges; this spontaneous formation process of ITB is insensitive to initial conditions, therefore, the ITB is a self-organized structure\cite{BakPRA88,KishimotoPoP96}. Figs. 3(f-h) demonstrate that the typical ITB expansion (at $r=0.28a$) is induced by the radially outward propagated avalanche, which starts from around $r=0.20a$ [Fig. 3(f)] when the gradient inside the ITB is raised by heating to a threshold value [Fig. 3(g)]. Fourier analysis shows that the toroidal mode number of the dominant mode of the burst in the outer region is different from in the inner region, when the avalanche propagates outward [Fig. 3(f)]. 
The S-curve for $r=0.20a$ [Fig. 3(i)] starts from a H-state followed by a transient L-state but quickly returns to the H-state after the avalanche burst; this demonstrates that the ITB is a self-organized structure that is robust\cite{BakPRA88} or resilient\cite{KishimotoPoP96} to perturbations.

The changes of $E_r$ structure before transition have been clearly demonstrated in Figs. 3(c, h). Figs. 3(e, j) indicate that both pressure gradient change\cite{WangPoP17} and poloidal flow change\cite{DiamondPRL94} contribute significantly to the $E_r$ change. The ion poloidal flow $u_{\theta}$ is calculated from the ion radial force balance equation,  
$E_r+u_{\theta}B_T-u_{\zeta}B_{P}- p'_i/(n_ie_i)=0$, 
with the toroidal flow $u_{\zeta}$, and $E_r$ directly given by the simulation results; $p_i=n_iT_i$; $B_{P}$ is the poloidal magnetic field.
Figs. 3(e, j) [Figs. 3(a, f)] show that the $E_r$ shear is significantly enhanced [the turbulence is significantly reduced] across the transition within the mesoscale region labeled by the two vertical [horizontal] black dashed lines. 

Since the shearing $E_r$ structure or ZF is a stabilizing (organizing) force while $T_i'$ is a driving (dissipating) force of the system, the picture of the ITB expansion revealed here can be summarized as follows. When the external heating raises $T_i'$  above a threshold value inside the ITB, a catastrophic burst is excited there; this burst propagates radially outward in avalanche, and induces an outward mesoscale expansion of the $E_r$ structure through nonlinearly excited ZFs; therefore the structure of the stabilizing force is expanded, and hence the ITB, a self-organized structure, is expanded by the avalanche. 

In summary, we have proposed the NEU method, which avoids the secularity problem in the perturbative ($\delta f$) computation, and significantly speeds up the computation by more than 10 times in a long-time nonlinear GK global simulation. Based on this critical progress, we have successfully revealed for the first time the formation dynamics of the ITB. We found that the emergence of the ITB is due to the inward propagated avalanche; the ITB is a self-organized structure and its outward expansion is the catastrophe induced by the outward propagated avalanche.  

The results may also add insight into the physics of edge transport barrier in the H-mode plasmas\cite{WagnerPPCF07}. Note that the NEU re-partitions the equilibrium and perturbation; the equilibrium, in addition to the perturbation, is evolved here by using the first-principle nonlinear GK simulation. This is different from Ref. \onlinecite{BarnesPoP10}, which evolves the equilibrium on the long-time scale by one-dimensional transport modeling with the flux-gradient relations extrapolated from the short-time GK simulation. 
 
\noindent{\bf Acknowledgments:}
S. W. thanks L. Chen and F. Zonca for critically reading the manuscript, W. X. Wang, J. Q. Li, Y. Xiao, for conversations on the precision of nonlinear GK simulations, and M. Xu, B. N. Wan, for conversations on the ITB experiments. T. W. thanks K. Imadera for communications on Ref. \onlinecite{ImaderaPPCF23} after this work is carried out. The authors are in debt to Z. Dai for valuable contributions to the implementation of heating and collision terms. This work was supported by the National MCF Energy R\&D Program of China under Grant No. 2019YFE03060000, and the National Natural Science Foundation of China under Grant No. 12075240. 


\begin{thebibliography}{58}%
\makeatletter
\providecommand \@ifxundefined [1]{%
 \@ifx{#1\undefined}
}%
\providecommand \@ifnum [1]{%
 \ifnum #1\expandafter \@firstoftwo
 \else \expandafter \@secondoftwo
 \fi
}%
\providecommand \@ifx [1]{%
 \ifx #1\expandafter \@firstoftwo
 \else \expandafter \@secondoftwo
 \fi
}%
\providecommand \natexlab [1]{#1}%
\providecommand \enquote  [1]{``#1''}%
\providecommand \bibnamefont  [1]{#1}%
\providecommand \bibfnamefont [1]{#1}%
\providecommand \citenamefont [1]{#1}%
\providecommand \href@noop [0]{\@secondoftwo}%
\providecommand \href [0]{\begingroup \@sanitize@url \@href}%
\providecommand \@href[1]{\@@startlink{#1}\@@href}%
\providecommand \@@href[1]{\endgroup#1\@@endlink}%
\providecommand \@sanitize@url [0]{\catcode `\\12\catcode `\$12\catcode
  `\&12\catcode `\#12\catcode `\^12\catcode `\_12\catcode `\%12\relax}%
\providecommand \@@startlink[1]{}%
\providecommand \@@endlink[0]{}%
\providecommand \url  [0]{\begingroup\@sanitize@url \@url }%
\providecommand \@url [1]{\endgroup\@href {#1}{\urlprefix }}%
\providecommand \urlprefix  [0]{URL }%
\providecommand \Eprint [0]{\href }%
\providecommand \doibase [0]{http://dx.doi.org/}%
\providecommand \selectlanguage [0]{\@gobble}%
\providecommand \bibinfo  [0]{\@secondoftwo}%
\providecommand \bibfield  [0]{\@secondoftwo}%
\providecommand \translation [1]{[#1]}%
\providecommand \BibitemOpen [0]{}%
\providecommand \bibitemStop [0]{}%
\providecommand \bibitemNoStop [0]{.\EOS\space}%
\providecommand \EOS [0]{\spacefactor3000\relax}%
\providecommand \BibitemShut  [1]{\csname bibitem#1\endcsname}%
\let\auto@bib@innerbib\@empty
\bibitem [{\citenamefont {Bak}\ \emph {et~al.}(1987)\citenamefont {Bak},
  \citenamefont {Tang},\ and\ \citenamefont {Wiesenfeld}}]{BakPRL87}%
  \BibitemOpen
  \bibfield  {author} {\bibinfo {author} {\bibfnamefont {P.}~\bibnamefont
  {Bak}}, \bibinfo {author} {\bibfnamefont {C.}~\bibnamefont {Tang}}, \ and\
  \bibinfo {author} {\bibfnamefont {K.}~\bibnamefont {Wiesenfeld}},\
  }\href@noop {} {\bibfield  {journal} {\bibinfo  {journal} {Phys. Rev. Lett.}\
  }\textbf {\bibinfo {volume} {59}},\ \bibinfo {pages} {381} (\bibinfo {year}
  {1987})}\BibitemShut {NoStop}%
\bibitem [{\citenamefont {Bak}\ \emph {et~al.}(1988)\citenamefont {Bak},
  \citenamefont {Tang},\ and\ \citenamefont {Wiesenfeld}}]{BakPRA88}%
  \BibitemOpen
  \bibfield  {author} {\bibinfo {author} {\bibfnamefont {P.}~\bibnamefont
  {Bak}}, \bibinfo {author} {\bibfnamefont {C.}~\bibnamefont {Tang}}, \ and\
  \bibinfo {author} {\bibfnamefont {K.}~\bibnamefont {Wiesenfeld}},\
  }\href@noop {} {\bibfield  {journal} {\bibinfo  {journal} {Phys. Rev. A}\
  }\textbf {\bibinfo {volume} {38}},\ \bibinfo {pages} {364} (\bibinfo {year}
  {1988})}\BibitemShut {NoStop}%
\bibitem [{\citenamefont {Sole}\ and\ \citenamefont
  {Manrubia}(1995)}]{SoleJTB95}%
  \BibitemOpen
  \bibfield  {author} {\bibinfo {author} {\bibfnamefont {R.~V.}\ \bibnamefont
  {Sole}}\ and\ \bibinfo {author} {\bibfnamefont {S.~C.}\ \bibnamefont
  {Manrubia}},\ }\href@noop {} {\bibfield  {journal} {\bibinfo  {journal} {J.
  Theor. Biol.}\ }\textbf {\bibinfo {volume} {173}},\ \bibinfo {pages} {31}
  (\bibinfo {year} {1995})}\BibitemShut {NoStop}%
\bibitem [{\citenamefont {Sornette}\ \emph {et~al.}(1990)\citenamefont
  {Sornette}, \citenamefont {Davy},\ and\ \citenamefont
  {Sornette}}]{SornetteJGR90}%
  \BibitemOpen
  \bibfield  {author} {\bibinfo {author} {\bibfnamefont {D.}~\bibnamefont
  {Sornette}}, \bibinfo {author} {\bibfnamefont {P.}~\bibnamefont {Davy}}, \
  and\ \bibinfo {author} {\bibfnamefont {A.}~\bibnamefont {Sornette}},\
  }\href@noop {} {\bibfield  {journal} {\bibinfo  {journal} {J. Geophys. Res.}\
  }\textbf {\bibinfo {volume} {95}},\ \bibinfo {pages} {17353} (\bibinfo {year}
  {1990})}\BibitemShut {NoStop}%
\bibitem [{\citenamefont {Nayak}\ and\ \citenamefont
  {Gupta}(2010)}]{NayakOLE10}%
  \BibitemOpen
  \bibfield  {author} {\bibinfo {author} {\bibfnamefont {B.~K.}\ \bibnamefont
  {Nayak}}\ and\ \bibinfo {author} {\bibfnamefont {M.~C.}\ \bibnamefont
  {Gupta}},\ }\href@noop {} {\bibfield  {journal} {\bibinfo  {journal} {Opt.
  Lasers Eng.}\ }\textbf {\bibinfo {volume} {48}},\ \bibinfo {pages} {940}
  (\bibinfo {year} {2010})}\BibitemShut {NoStop}%
\bibitem [{\citenamefont {McDevitt}\ \emph {et~al.}(2008)\citenamefont
  {McDevitt}, \citenamefont {Tang},\ and\ \citenamefont {Guo}}]{MacakJEC08}%
  \BibitemOpen
  \bibfield  {author} {\bibinfo {author} {\bibfnamefont {C.~J.}\ \bibnamefont
  {McDevitt}}, \bibinfo {author} {\bibfnamefont {X.}~\bibnamefont {Tang}}, \
  and\ \bibinfo {author} {\bibfnamefont {Z.}~\bibnamefont {Guo}},\ }\href@noop
  {} {\bibfield  {journal} {\bibinfo  {journal} {J. Elect. Chemistry}\ }\textbf
  {\bibinfo {volume} {621}},\ \bibinfo {pages} {254} (\bibinfo {year}
  {2008})}\BibitemShut {NoStop}%
\bibitem [{\citenamefont {Kishimoto}\ \emph {et~al.}(1996)\citenamefont
  {Kishimoto}, \citenamefont {Tajima}, \citenamefont {Horton}, \citenamefont
  {LeBrun},\ and\ \citenamefont {Kim}}]{KishimotoPoP96}%
  \BibitemOpen
  \bibfield  {author} {\bibinfo {author} {\bibfnamefont {Y.}~\bibnamefont
  {Kishimoto}}, \bibinfo {author} {\bibfnamefont {T.}~\bibnamefont {Tajima}},
  \bibinfo {author} {\bibfnamefont {W.}~\bibnamefont {Horton}}, \bibinfo
  {author} {\bibfnamefont {M.~J.}\ \bibnamefont {LeBrun}}, \ and\ \bibinfo
  {author} {\bibfnamefont {J.~Y.}\ \bibnamefont {Kim}},\ }\href@noop {}
  {\bibfield  {journal} {\bibinfo  {journal} {Phys. Plasmas}\ }\textbf
  {\bibinfo {volume} {3}},\ \bibinfo {pages} {1289} (\bibinfo {year}
  {1996})}\BibitemShut {NoStop}%
\bibitem [{\citenamefont {Shimada}\ \emph {et~al.}(2007)\citenamefont
  {Shimada}, \citenamefont {Campbell}, \citenamefont {Mukhovatov},
  \citenamefont {Fujiwara}, \citenamefont {Kirneva}, \citenamefont {Lackner},
  \citenamefont {Nagami}, \citenamefont {Pustovitov}, \citenamefont {Uckan},\
  and\ \citenamefont {Wesley}}]{ShimadaNF07}%
  \BibitemOpen
  \bibfield  {author} {\bibinfo {author} {\bibfnamefont {M.}~\bibnamefont
  {Shimada}}, \bibinfo {author} {\bibfnamefont {D.~J.}\ \bibnamefont
  {Campbell}}, \bibinfo {author} {\bibfnamefont {V.}~\bibnamefont
  {Mukhovatov}}, \bibinfo {author} {\bibfnamefont {M.}~\bibnamefont
  {Fujiwara}}, \bibinfo {author} {\bibfnamefont {N.}~\bibnamefont {Kirneva}},
  \bibinfo {author} {\bibfnamefont {K.}~\bibnamefont {Lackner}}, \bibinfo
  {author} {\bibfnamefont {M.}~\bibnamefont {Nagami}}, \bibinfo {author}
  {\bibfnamefont {V.~D.}\ \bibnamefont {Pustovitov}}, \bibinfo {author}
  {\bibfnamefont {N.}~\bibnamefont {Uckan}}, \ and\ \bibinfo {author}
  {\bibfnamefont {J.}~\bibnamefont {Wesley}},\ }\href@noop {} {\bibfield
  {journal} {\bibinfo  {journal} {Nucl. Fusion}\ }\textbf {\bibinfo {volume}
  {47}},\ \bibinfo {pages} {S1} (\bibinfo {year} {2007})}\BibitemShut {NoStop}%
\bibitem [{\citenamefont {Ongena}\ \emph {et~al.}(2016)\citenamefont {Ongena},
  \citenamefont {Wolf},\ and\ \citenamefont {Zohm}}]{OngenaNP16}%
  \BibitemOpen
  \bibfield  {author} {\bibinfo {author} {\bibfnamefont {J.}~\bibnamefont
  {Ongena}}, \bibinfo {author} {\bibfnamefont {R.}~\bibnamefont {Wolf}}, \ and\
  \bibinfo {author} {\bibfnamefont {H.}~\bibnamefont {Zohm}},\ }\href@noop {}
  {\bibfield  {journal} {\bibinfo  {journal} {Nat. Phys.}\ }\textbf {\bibinfo
  {volume} {12}},\ \bibinfo {pages} {398} (\bibinfo {year} {2016})}\BibitemShut
  {NoStop}%
\bibitem [{\citenamefont {Fasoli}(2023)}]{FasoliPRL23}%
  \BibitemOpen
  \bibfield  {author} {\bibinfo {author} {\bibfnamefont {A.}~\bibnamefont
  {Fasoli}},\ }\href@noop {} {\bibfield  {journal} {\bibinfo  {journal} {Phys.
  Rev. Lett.}\ }\textbf {\bibinfo {volume} {130}},\ \bibinfo {pages} {220001}
  (\bibinfo {year} {2023})}\BibitemShut {NoStop}%
\bibitem [{\citenamefont {Doyle}\ \emph {et~al.}(2007)\citenamefont {Doyle},
  \citenamefont {Houlberg}, \citenamefont {Kamada}, \citenamefont {Mukhovatov},
  \citenamefont {Osborne},\ and\ \citenamefont {et. al.}}]{DoyleNF07}%
  \BibitemOpen
  \bibfield  {author} {\bibinfo {author} {\bibfnamefont {E.~J.}\ \bibnamefont
  {Doyle}}, \bibinfo {author} {\bibfnamefont {W.~A.}\ \bibnamefont {Houlberg}},
  \bibinfo {author} {\bibfnamefont {Y.}~\bibnamefont {Kamada}}, \bibinfo
  {author} {\bibfnamefont {V.}~\bibnamefont {Mukhovatov}}, \bibinfo {author}
  {\bibfnamefont {T.~H.}\ \bibnamefont {Osborne}}, \ and\ \bibinfo {author}
  {\bibnamefont {et. al.}},\ }\href@noop {} {\bibfield  {journal} {\bibinfo
  {journal} {Nucl. Fusion}\ }\textbf {\bibinfo {volume} {47}},\ \bibinfo
  {pages} {S18} (\bibinfo {year} {2007})}\BibitemShut {NoStop}%
\bibitem [{\citenamefont {Koide}\ \emph {et~al.}(1994)\citenamefont {Koide},
  \citenamefont {Kikuchi}, \citenamefont {Mori}, \citenamefont {Tsuji},
  \citenamefont {Ishida}, \citenamefont {Asakura}, \citenamefont {Kamada},
  \citenamefont {Nishitani}, \citenamefont {Kawano}, \citenamefont {Hatae},
  \citenamefont {Fujita}, \citenamefont {Fukuda}, \citenamefont {Sakasai},
  \citenamefont {Yoshino},\ and\ \citenamefont {Neyatani}}]{Koide-JT60U-PRL94}%
  \BibitemOpen
  \bibfield  {author} {\bibinfo {author} {\bibfnamefont {Y.}~\bibnamefont
  {Koide}}, \bibinfo {author} {\bibfnamefont {M.}~\bibnamefont {Kikuchi}},
  \bibinfo {author} {\bibfnamefont {M.}~\bibnamefont {Mori}}, \bibinfo {author}
  {\bibfnamefont {S.}~\bibnamefont {Tsuji}}, \bibinfo {author} {\bibfnamefont
  {S.}~\bibnamefont {Ishida}}, \bibinfo {author} {\bibfnamefont
  {N.}~\bibnamefont {Asakura}}, \bibinfo {author} {\bibfnamefont
  {Y.}~\bibnamefont {Kamada}}, \bibinfo {author} {\bibfnamefont
  {T.}~\bibnamefont {Nishitani}}, \bibinfo {author} {\bibfnamefont
  {Y.}~\bibnamefont {Kawano}}, \bibinfo {author} {\bibfnamefont
  {T.}~\bibnamefont {Hatae}}, \bibinfo {author} {\bibfnamefont
  {T.}~\bibnamefont {Fujita}}, \bibinfo {author} {\bibfnamefont
  {T.}~\bibnamefont {Fukuda}}, \bibinfo {author} {\bibfnamefont
  {A.}~\bibnamefont {Sakasai}}, \bibinfo {author} {\bibfnamefont
  {R.}~\bibnamefont {Yoshino}}, \ and\ \bibinfo {author} {\bibfnamefont
  {Y.}~\bibnamefont {Neyatani}},\ }\href@noop {} {\bibfield  {journal}
  {\bibinfo  {journal} {Phys. Rev. Lett.}\ }\textbf {\bibinfo {volume} {72}},\
  \bibinfo {pages} {3662} (\bibinfo {year} {1994})}\BibitemShut {NoStop}%
\bibitem [{\citenamefont {Fujita}\ \emph {et~al.}(2001)\citenamefont {Fujita},
  \citenamefont {Oikawa}, \citenamefont {Suzuki}, \citenamefont {Ide},
  \citenamefont {Sakamoto}, \citenamefont {Koide}, \citenamefont {Hatae},
  \citenamefont {Naito}, \citenamefont {Isayama}, \citenamefont {Hayashi},\
  and\ \citenamefont {Shirai}}]{Fujita-JT60U-PRL01}%
  \BibitemOpen
  \bibfield  {author} {\bibinfo {author} {\bibfnamefont {T.}~\bibnamefont
  {Fujita}}, \bibinfo {author} {\bibfnamefont {T.}~\bibnamefont {Oikawa}},
  \bibinfo {author} {\bibfnamefont {T.}~\bibnamefont {Suzuki}}, \bibinfo
  {author} {\bibfnamefont {S.}~\bibnamefont {Ide}}, \bibinfo {author}
  {\bibfnamefont {Y.}~\bibnamefont {Sakamoto}}, \bibinfo {author}
  {\bibfnamefont {Y.}~\bibnamefont {Koide}}, \bibinfo {author} {\bibfnamefont
  {T.}~\bibnamefont {Hatae}}, \bibinfo {author} {\bibfnamefont
  {O.}~\bibnamefont {Naito}}, \bibinfo {author} {\bibfnamefont
  {A.}~\bibnamefont {Isayama}}, \bibinfo {author} {\bibfnamefont
  {N.}~\bibnamefont {Hayashi}}, \ and\ \bibinfo {author} {\bibfnamefont
  {H.}~\bibnamefont {Shirai}},\ }\href@noop {} {\bibfield  {journal} {\bibinfo
  {journal} {Phys. Rev. Lett.}\ }\textbf {\bibinfo {volume} {87}},\ \bibinfo
  {pages} {245001} (\bibinfo {year} {2001})}\BibitemShut {NoStop}%
\bibitem [{\citenamefont {Levinton}\ \emph {et~al.}(1995)\citenamefont
  {Levinton}, \citenamefont {Zarnstorff}, \citenamefont {Batha}, \citenamefont
  {Bell}, \citenamefont {Bell}, \citenamefont {Budny}, \citenamefont {Bush},
  \citenamefont {Chang}, \citenamefont {Fredrickson}, \citenamefont {Janos},
  \citenamefont {Manickam}, \citenamefont {Ramsey}, \citenamefont {Sabbagh},
  \citenamefont {Schmidt}, \citenamefont {Synakowski},\ and\ \citenamefont
  {Taylor}}]{Levinton-TFTR-PRL95}%
  \BibitemOpen
  \bibfield  {author} {\bibinfo {author} {\bibfnamefont {F.~M.}\ \bibnamefont
  {Levinton}}, \bibinfo {author} {\bibfnamefont {M.~C.}\ \bibnamefont
  {Zarnstorff}}, \bibinfo {author} {\bibfnamefont {S.~H.}\ \bibnamefont
  {Batha}}, \bibinfo {author} {\bibfnamefont {M.}~\bibnamefont {Bell}},
  \bibinfo {author} {\bibfnamefont {R.~E.}\ \bibnamefont {Bell}}, \bibinfo
  {author} {\bibfnamefont {R.~V.}\ \bibnamefont {Budny}}, \bibinfo {author}
  {\bibfnamefont {C.}~\bibnamefont {Bush}}, \bibinfo {author} {\bibfnamefont
  {Z.}~\bibnamefont {Chang}}, \bibinfo {author} {\bibfnamefont
  {E.}~\bibnamefont {Fredrickson}}, \bibinfo {author} {\bibfnamefont
  {A.}~\bibnamefont {Janos}}, \bibinfo {author} {\bibfnamefont
  {J.}~\bibnamefont {Manickam}}, \bibinfo {author} {\bibfnamefont
  {A.}~\bibnamefont {Ramsey}}, \bibinfo {author} {\bibfnamefont {S.~A.}\
  \bibnamefont {Sabbagh}}, \bibinfo {author} {\bibfnamefont {G.~L.}\
  \bibnamefont {Schmidt}}, \bibinfo {author} {\bibfnamefont {E.~J.}\
  \bibnamefont {Synakowski}}, \ and\ \bibinfo {author} {\bibfnamefont
  {G.}~\bibnamefont {Taylor}},\ }\href@noop {} {\bibfield  {journal} {\bibinfo
  {journal} {Phys. Rev. Lett.}\ }\textbf {\bibinfo {volume} {75}},\ \bibinfo
  {pages} {4417} (\bibinfo {year} {1995})}\BibitemShut {NoStop}%
\bibitem [{\citenamefont {Synakowski}\ \emph {et~al.}(1997)\citenamefont
  {Synakowski}, \citenamefont {Batha}, \citenamefont {Beer}, \citenamefont
  {Bell}, \citenamefont {Bell}, \citenamefont {Budny}, \citenamefont {Bush},
  \citenamefont {Efthimion}, \citenamefont {Hammett}, \citenamefont {Hahm},
  \citenamefont {LeBlanc}, \citenamefont {Levinton}, \citenamefont {Mazzucato},
  \citenamefont {Park}, \citenamefont {Ramsey}, \citenamefont {Rewoldt},
  \citenamefont {Scott}, \citenamefont {Schmidt}, \citenamefont {Tang},
  \citenamefont {Taylor},\ and\ \citenamefont {Zarnstorff}}]{SynakowskiPRL97}%
  \BibitemOpen
  \bibfield  {author} {\bibinfo {author} {\bibfnamefont {E.~J.}\ \bibnamefont
  {Synakowski}}, \bibinfo {author} {\bibfnamefont {S.~H.}\ \bibnamefont
  {Batha}}, \bibinfo {author} {\bibfnamefont {M.~A.}\ \bibnamefont {Beer}},
  \bibinfo {author} {\bibfnamefont {M.~G.}\ \bibnamefont {Bell}}, \bibinfo
  {author} {\bibfnamefont {R.~E.}\ \bibnamefont {Bell}}, \bibinfo {author}
  {\bibfnamefont {R.~V.}\ \bibnamefont {Budny}}, \bibinfo {author}
  {\bibfnamefont {C.~E.}\ \bibnamefont {Bush}}, \bibinfo {author}
  {\bibfnamefont {P.~C.}\ \bibnamefont {Efthimion}}, \bibinfo {author}
  {\bibfnamefont {G.~W.}\ \bibnamefont {Hammett}}, \bibinfo {author}
  {\bibfnamefont {T.~S.}\ \bibnamefont {Hahm}}, \bibinfo {author}
  {\bibfnamefont {B.}~\bibnamefont {LeBlanc}}, \bibinfo {author} {\bibfnamefont
  {F.}~\bibnamefont {Levinton}}, \bibinfo {author} {\bibfnamefont
  {E.}~\bibnamefont {Mazzucato}}, \bibinfo {author} {\bibfnamefont
  {H.}~\bibnamefont {Park}}, \bibinfo {author} {\bibfnamefont {A.~T.}\
  \bibnamefont {Ramsey}}, \bibinfo {author} {\bibfnamefont {G.}~\bibnamefont
  {Rewoldt}}, \bibinfo {author} {\bibfnamefont {S.~D.}\ \bibnamefont {Scott}},
  \bibinfo {author} {\bibfnamefont {G.}~\bibnamefont {Schmidt}}, \bibinfo
  {author} {\bibfnamefont {W.~M.}\ \bibnamefont {Tang}}, \bibinfo {author}
  {\bibfnamefont {G.}~\bibnamefont {Taylor}}, \ and\ \bibinfo {author}
  {\bibfnamefont {M.~C.}\ \bibnamefont {Zarnstorff}},\ }\href@noop {}
  {\bibfield  {journal} {\bibinfo  {journal} {Phys. Rev. Lett.}\ }\textbf
  {\bibinfo {volume} {78}},\ \bibinfo {pages} {2972} (\bibinfo {year}
  {1997})}\BibitemShut {NoStop}%
\bibitem [{\citenamefont {Strait}\ \emph {et~al.}(1995)\citenamefont {Strait},
  \citenamefont {Lao}, \citenamefont {Mauel}, \citenamefont {Rice},
  \citenamefont {Taylor}, \citenamefont {Burrell}, \citenamefont {Chu},
  \citenamefont {Lazarus}, \citenamefont {Osborne}, \citenamefont {Thompson},\
  and\ \citenamefont {Turnbull}}]{Strait-DIIID-PRL95}%
  \BibitemOpen
  \bibfield  {author} {\bibinfo {author} {\bibfnamefont {E.~J.}\ \bibnamefont
  {Strait}}, \bibinfo {author} {\bibfnamefont {L.~L.}\ \bibnamefont {Lao}},
  \bibinfo {author} {\bibfnamefont {M.~E.}\ \bibnamefont {Mauel}}, \bibinfo
  {author} {\bibfnamefont {B.~W.}\ \bibnamefont {Rice}}, \bibinfo {author}
  {\bibfnamefont {T.~S.}\ \bibnamefont {Taylor}}, \bibinfo {author}
  {\bibfnamefont {K.~H.}\ \bibnamefont {Burrell}}, \bibinfo {author}
  {\bibfnamefont {M.~S.}\ \bibnamefont {Chu}}, \bibinfo {author} {\bibfnamefont
  {E.~A.}\ \bibnamefont {Lazarus}}, \bibinfo {author} {\bibfnamefont {T.~H.}\
  \bibnamefont {Osborne}}, \bibinfo {author} {\bibfnamefont {S.~J.}\
  \bibnamefont {Thompson}}, \ and\ \bibinfo {author} {\bibfnamefont {A.~D.}\
  \bibnamefont {Turnbull}},\ }\href@noop {} {\bibfield  {journal} {\bibinfo
  {journal} {Phys. Rev. Lett.}\ }\textbf {\bibinfo {volume} {75}},\ \bibinfo
  {pages} {4421} (\bibinfo {year} {1995})}\BibitemShut {NoStop}%
\bibitem [{\citenamefont {Burrell}\ \emph {et~al.}(1998)\citenamefont
  {Burrell}, \citenamefont {Austin}, \citenamefont {Greefield}, \citenamefont
  {Lao}, \citenamefont {Rice}, \citenamefont {Staebler},\ and\ \citenamefont
  {Stallard}}]{BurrellPPCF98}%
  \BibitemOpen
  \bibfield  {author} {\bibinfo {author} {\bibfnamefont {K.~H.}\ \bibnamefont
  {Burrell}}, \bibinfo {author} {\bibfnamefont {M.~E.}\ \bibnamefont {Austin}},
  \bibinfo {author} {\bibfnamefont {C.~M.}\ \bibnamefont {Greefield}}, \bibinfo
  {author} {\bibfnamefont {L.~L.}\ \bibnamefont {Lao}}, \bibinfo {author}
  {\bibfnamefont {B.~W.}\ \bibnamefont {Rice}}, \bibinfo {author}
  {\bibfnamefont {G.~M.}\ \bibnamefont {Staebler}}, \ and\ \bibinfo {author}
  {\bibfnamefont {B.~W.}\ \bibnamefont {Stallard}},\ }\href@noop {} {\bibfield
  {journal} {\bibinfo  {journal} {Plasma Phys. Control. Fusion}\ }\textbf
  {\bibinfo {volume} {40}},\ \bibinfo {pages} {1585} (\bibinfo {year}
  {1998})}\BibitemShut {NoStop}%
\bibitem [{\citenamefont {Rettig}\ \emph {et~al.}(1998)\citenamefont {Rettig},
  \citenamefont {Burrell}, \citenamefont {Stallard}, \citenamefont {McKee},
  \citenamefont {Staebler}, \citenamefont {Rhodes}, \citenamefont
  {Greenfield},\ and\ \citenamefont {Peebles}}]{RettigPoP98}%
  \BibitemOpen
  \bibfield  {author} {\bibinfo {author} {\bibfnamefont {C.~L.}\ \bibnamefont
  {Rettig}}, \bibinfo {author} {\bibfnamefont {K.~H.}\ \bibnamefont {Burrell}},
  \bibinfo {author} {\bibfnamefont {B.~W.}\ \bibnamefont {Stallard}}, \bibinfo
  {author} {\bibfnamefont {G.~R.}\ \bibnamefont {McKee}}, \bibinfo {author}
  {\bibfnamefont {G.~M.}\ \bibnamefont {Staebler}}, \bibinfo {author}
  {\bibfnamefont {T.~L.}\ \bibnamefont {Rhodes}}, \bibinfo {author}
  {\bibfnamefont {C.~M.}\ \bibnamefont {Greenfield}}, \ and\ \bibinfo {author}
  {\bibfnamefont {W.~A.}\ \bibnamefont {Peebles}},\ }\href@noop {} {\bibfield
  {journal} {\bibinfo  {journal} {Phys. Plasmas}\ }\textbf {\bibinfo {volume}
  {5}},\ \bibinfo {pages} {1727} (\bibinfo {year} {1998})}\BibitemShut
  {NoStop}%
\bibitem [{\citenamefont {Crisanti}\ \emph {et~al.}(2002)\citenamefont
  {Crisanti}, \citenamefont {Litaudon}, \citenamefont {Mailloux}, \citenamefont
  {Mazon}, \citenamefont {Barbato}, \citenamefont {Baranov}, \citenamefont
  {Becoulet}, \citenamefont {Becoulet}, \citenamefont {Challis}, \citenamefont
  {Conway}, \citenamefont {Dux}, \citenamefont {Erikkkon}, \citenamefont
  {Esposito}, \citenamefont {Frigione}, \citenamefont {Hennequin},
  \citenamefont {Giroud}, \citenamefont {Hawkes}, \citenamefont {Huysmans},
  \citenamefont {Imbeaux}, \citenamefont {Joffrin}, \citenamefont {Lomas},
  \citenamefont {Lotte}, \citenamefont {Maget}, \citenamefont {Mantsinen},
  \citenamefont {Moreau}, \citenamefont {Rimini}, \citenamefont {Riva},
  \citenamefont {Sarazin}, \citenamefont {Tresset}, \citenamefont {Tuccillo},\
  and\ \citenamefont {Zastrow}}]{Crisanti-JET-PRL02}%
  \BibitemOpen
  \bibfield  {author} {\bibinfo {author} {\bibfnamefont {F.}~\bibnamefont
  {Crisanti}}, \bibinfo {author} {\bibfnamefont {X.}~\bibnamefont {Litaudon}},
  \bibinfo {author} {\bibfnamefont {J.}~\bibnamefont {Mailloux}}, \bibinfo
  {author} {\bibfnamefont {D.}~\bibnamefont {Mazon}}, \bibinfo {author}
  {\bibfnamefont {E.}~\bibnamefont {Barbato}}, \bibinfo {author} {\bibfnamefont
  {Y.}~\bibnamefont {Baranov}}, \bibinfo {author} {\bibfnamefont
  {A.}~\bibnamefont {Becoulet}}, \bibinfo {author} {\bibfnamefont
  {M.}~\bibnamefont {Becoulet}}, \bibinfo {author} {\bibfnamefont {C.~D.}\
  \bibnamefont {Challis}}, \bibinfo {author} {\bibfnamefont {G.~D.}\
  \bibnamefont {Conway}}, \bibinfo {author} {\bibfnamefont {R.}~\bibnamefont
  {Dux}}, \bibinfo {author} {\bibfnamefont {L.~G.}\ \bibnamefont {Erikkkon}},
  \bibinfo {author} {\bibfnamefont {B.}~\bibnamefont {Esposito}}, \bibinfo
  {author} {\bibfnamefont {D.}~\bibnamefont {Frigione}}, \bibinfo {author}
  {\bibfnamefont {P.}~\bibnamefont {Hennequin}}, \bibinfo {author}
  {\bibfnamefont {C.}~\bibnamefont {Giroud}}, \bibinfo {author} {\bibfnamefont
  {N.}~\bibnamefont {Hawkes}}, \bibinfo {author} {\bibfnamefont
  {G.}~\bibnamefont {Huysmans}}, \bibinfo {author} {\bibfnamefont
  {F.}~\bibnamefont {Imbeaux}}, \bibinfo {author} {\bibfnamefont
  {E.}~\bibnamefont {Joffrin}}, \bibinfo {author} {\bibfnamefont
  {P.}~\bibnamefont {Lomas}}, \bibinfo {author} {\bibfnamefont
  {P.}~\bibnamefont {Lotte}}, \bibinfo {author} {\bibfnamefont
  {P.}~\bibnamefont {Maget}}, \bibinfo {author} {\bibfnamefont
  {M.}~\bibnamefont {Mantsinen}}, \bibinfo {author} {\bibfnamefont
  {D.}~\bibnamefont {Moreau}}, \bibinfo {author} {\bibfnamefont
  {F.}~\bibnamefont {Rimini}}, \bibinfo {author} {\bibfnamefont
  {M.}~\bibnamefont {Riva}}, \bibinfo {author} {\bibfnamefont {Y.}~\bibnamefont
  {Sarazin}}, \bibinfo {author} {\bibfnamefont {G.}~\bibnamefont {Tresset}},
  \bibinfo {author} {\bibfnamefont {A.~A.}\ \bibnamefont {Tuccillo}}, \ and\
  \bibinfo {author} {\bibfnamefont {K.~D.}\ \bibnamefont {Zastrow}},\
  }\href@noop {} {\bibfield  {journal} {\bibinfo  {journal} {Phys. Rev. Lett.}\
  }\textbf {\bibinfo {volume} {88}},\ \bibinfo {pages} {145004} (\bibinfo
  {year} {2002})}\BibitemShut {NoStop}%
\bibitem [{\citenamefont {Mazzi}\ \emph {et~al.}(2022)\citenamefont {Mazzi},
  \citenamefont {J.}, \citenamefont {Zarzoso}, \citenamefont {Kazakov},
  \citenamefont {Ongena}, \citenamefont {Dreval}, \citenamefont {M.},
  \citenamefont {Stancar}, \citenamefont {Szepesi}, \citenamefont {Eriksson},
  \citenamefont {Sahlberg}, \citenamefont {Benkadda},\ and\ \citenamefont
  {Contributors}}]{MazziNP22}%
  \BibitemOpen
  \bibfield  {author} {\bibinfo {author} {\bibfnamefont {S.}~\bibnamefont
  {Mazzi}}, \bibinfo {author} {\bibfnamefont {G.}~\bibnamefont {J.}}, \bibinfo
  {author} {\bibfnamefont {D.}~\bibnamefont {Zarzoso}}, \bibinfo {author}
  {\bibfnamefont {Y.~O.}\ \bibnamefont {Kazakov}}, \bibinfo {author}
  {\bibfnamefont {J.}~\bibnamefont {Ongena}}, \bibinfo {author} {\bibfnamefont
  {M.}~\bibnamefont {Dreval}}, \bibinfo {author} {\bibfnamefont
  {N.}~\bibnamefont {M.}}, \bibinfo {author} {\bibfnamefont {Z.}~\bibnamefont
  {Stancar}}, \bibinfo {author} {\bibfnamefont {G.}~\bibnamefont {Szepesi}},
  \bibinfo {author} {\bibfnamefont {J.}~\bibnamefont {Eriksson}}, \bibinfo
  {author} {\bibfnamefont {A.}~\bibnamefont {Sahlberg}}, \bibinfo {author}
  {\bibfnamefont {S.}~\bibnamefont {Benkadda}}, \ and\ \bibinfo {author}
  {\bibfnamefont {J.}~\bibnamefont {Contributors}},\ }\href@noop {} {\bibfield
  {journal} {\bibinfo  {journal} {Nat. Phys.}\ }\textbf {\bibinfo {volume}
  {18}},\ \bibinfo {pages} {776} (\bibinfo {year} {2022})}\BibitemShut
  {NoStop}%
\bibitem [{\citenamefont {Hobirk}\ \emph {et~al.}(2001)\citenamefont {Hobirk},
  \citenamefont {Wolf}, \citenamefont {Gruber}, \citenamefont {Gude},
  \citenamefont {Gunter}, \citenamefont {Kurzan}, \citenamefont {Maraschek},
  \citenamefont {McCarthy}, \citenamefont {Meister}, \citenamefont {Peeters},
  \citenamefont {Pereverzev}, \citenamefont {Stober}, \citenamefont
  {Trutterer},\ and\ \citenamefont {Team}}]{Hobirk-ASDEX-PRL01}%
  \BibitemOpen
  \bibfield  {author} {\bibinfo {author} {\bibfnamefont {J.}~\bibnamefont
  {Hobirk}}, \bibinfo {author} {\bibfnamefont {R.~C.}\ \bibnamefont {Wolf}},
  \bibinfo {author} {\bibfnamefont {O.}~\bibnamefont {Gruber}}, \bibinfo
  {author} {\bibfnamefont {A.}~\bibnamefont {Gude}}, \bibinfo {author}
  {\bibfnamefont {S.}~\bibnamefont {Gunter}}, \bibinfo {author} {\bibfnamefont
  {B.}~\bibnamefont {Kurzan}}, \bibinfo {author} {\bibfnamefont
  {M.}~\bibnamefont {Maraschek}}, \bibinfo {author} {\bibfnamefont {P.~J.}\
  \bibnamefont {McCarthy}}, \bibinfo {author} {\bibfnamefont {H.}~\bibnamefont
  {Meister}}, \bibinfo {author} {\bibfnamefont {A.~G.}\ \bibnamefont
  {Peeters}}, \bibinfo {author} {\bibfnamefont {G.~V.}\ \bibnamefont
  {Pereverzev}}, \bibinfo {author} {\bibfnamefont {J.}~\bibnamefont {Stober}},
  \bibinfo {author} {\bibfnamefont {W.}~\bibnamefont {Trutterer}}, \ and\
  \bibinfo {author} {\bibfnamefont {A.~U.}\ \bibnamefont {Team}},\ }\href@noop
  {} {\bibfield  {journal} {\bibinfo  {journal} {Phys. Rev. Lett.}\ }\textbf
  {\bibinfo {volume} {87}},\ \bibinfo {pages} {085002} (\bibinfo {year}
  {2001})}\BibitemShut {NoStop}%
\bibitem [{\citenamefont {Yu}\ \emph {et~al.}(2016)\citenamefont {Yu},
  \citenamefont {Wei}, \citenamefont {Liu}, \citenamefont {Dong}, \citenamefont
  {Ida}, \citenamefont {Itoh}, \citenamefont {Sun}, \citenamefont {Cao},
  \citenamefont {Shi}, \citenamefont {Wang}, \citenamefont {Xiao},
  \citenamefont {Yuan}, \citenamefont {Du}, \citenamefont {He}, \citenamefont
  {Chen}, \citenamefont {Ma}, \citenamefont {Itoh}, \citenamefont {Zhao},
  \citenamefont {Y.}, \citenamefont {Wang}, \citenamefont {Ji}, \citenamefont
  {Zhong}, \citenamefont {Li}, \citenamefont {Gao}, \citenamefont {Deng},
  \citenamefont {Liu}, \citenamefont {Xu}, \citenamefont {Yan}, \citenamefont
  {Yang}, \citenamefont {Ding}, \citenamefont {Duan}, \citenamefont {Liu},\
  and\ \citenamefont {Team}}]{Yu-HL2A-NF16}%
  \BibitemOpen
  \bibfield  {author} {\bibinfo {author} {\bibfnamefont {D.~L.}\ \bibnamefont
  {Yu}}, \bibinfo {author} {\bibfnamefont {Y.~L.}\ \bibnamefont {Wei}},
  \bibinfo {author} {\bibfnamefont {L.}~\bibnamefont {Liu}}, \bibinfo {author}
  {\bibfnamefont {J.~Q.}\ \bibnamefont {Dong}}, \bibinfo {author}
  {\bibfnamefont {K.}~\bibnamefont {Ida}}, \bibinfo {author} {\bibfnamefont
  {K.}~\bibnamefont {Itoh}}, \bibinfo {author} {\bibfnamefont {A.~P.}\
  \bibnamefont {Sun}}, \bibinfo {author} {\bibfnamefont {J.~Y.}\ \bibnamefont
  {Cao}}, \bibinfo {author} {\bibfnamefont {Z.~B.}\ \bibnamefont {Shi}},
  \bibinfo {author} {\bibfnamefont {Z.~X.}\ \bibnamefont {Wang}}, \bibinfo
  {author} {\bibfnamefont {Y.}~\bibnamefont {Xiao}}, \bibinfo {author}
  {\bibfnamefont {B.~S.}\ \bibnamefont {Yuan}}, \bibinfo {author}
  {\bibfnamefont {H.~R.}\ \bibnamefont {Du}}, \bibinfo {author} {\bibfnamefont
  {X.~X.}\ \bibnamefont {He}}, \bibinfo {author} {\bibfnamefont {W.~J.}\
  \bibnamefont {Chen}}, \bibinfo {author} {\bibfnamefont {Q.}~\bibnamefont
  {Ma}}, \bibinfo {author} {\bibfnamefont {S.~I.}\ \bibnamefont {Itoh}},
  \bibinfo {author} {\bibfnamefont {K.~J.}\ \bibnamefont {Zhao}}, \bibinfo
  {author} {\bibfnamefont {Z.}~\bibnamefont {Y.}}, \bibinfo {author}
  {\bibfnamefont {J.}~\bibnamefont {Wang}}, \bibinfo {author} {\bibfnamefont
  {X.~Q.}\ \bibnamefont {Ji}}, \bibinfo {author} {\bibfnamefont {W.~L.}\
  \bibnamefont {Zhong}}, \bibinfo {author} {\bibfnamefont {Y.~G.}\ \bibnamefont
  {Li}}, \bibinfo {author} {\bibfnamefont {J.~M.}\ \bibnamefont {Gao}},
  \bibinfo {author} {\bibfnamefont {W.}~\bibnamefont {Deng}}, \bibinfo {author}
  {\bibfnamefont {Y.}~\bibnamefont {Liu}}, \bibinfo {author} {\bibfnamefont
  {Y.}~\bibnamefont {Xu}}, \bibinfo {author} {\bibfnamefont {L.~W.}\
  \bibnamefont {Yan}}, \bibinfo {author} {\bibfnamefont {Q.~W.}\ \bibnamefont
  {Yang}}, \bibinfo {author} {\bibfnamefont {X.~T.}\ \bibnamefont {Ding}},
  \bibinfo {author} {\bibfnamefont {X.~R.}\ \bibnamefont {Duan}}, \bibinfo
  {author} {\bibfnamefont {Y.}~\bibnamefont {Liu}}, \ and\ \bibinfo {author}
  {\bibfnamefont {H.-A.}\ \bibnamefont {Team}},\ }\href@noop {} {\bibfield
  {journal} {\bibinfo  {journal} {Nucl. Fusion}\ }\textbf {\bibinfo {volume}
  {56}},\ \bibinfo {pages} {056003} (\bibinfo {year} {2016})}\BibitemShut
  {NoStop}%
\bibitem [{\citenamefont {Li}\ \emph {et~al.}(2022)\citenamefont {Li},
  \citenamefont {Zou}, \citenamefont {Xu}, \citenamefont {Chu}, \citenamefont
  {Feng}, \citenamefont {Lian}, \citenamefont {Liu}, \citenamefont {Liu},
  \citenamefont {Han}, \citenamefont {Dong}, \citenamefont {Wang},
  \citenamefont {Liu}, \citenamefont {Zang}, \citenamefont {Wang},
  \citenamefont {Zhou}, \citenamefont {Huang}, \citenamefont {Hu},
  \citenamefont {Zhou}, \citenamefont {Qu}, \citenamefont {Chen}, \citenamefont
  {Lin}, \citenamefont {Zhang}, \citenamefont {Qian}, \citenamefont {Hu},
  \citenamefont {Xu}, \citenamefont {Chen}, \citenamefont {Lu}, \citenamefont
  {Liu}, \citenamefont {Song}, \citenamefont {Li},\ and\ \citenamefont
  {Gong}}]{LiPRL22}%
  \BibitemOpen
  \bibfield  {author} {\bibinfo {author} {\bibfnamefont {E.}~\bibnamefont
  {Li}}, \bibinfo {author} {\bibfnamefont {X.~L.}\ \bibnamefont {Zou}},
  \bibinfo {author} {\bibfnamefont {L.~Q.}\ \bibnamefont {Xu}}, \bibinfo
  {author} {\bibfnamefont {Y.~Q.}\ \bibnamefont {Chu}}, \bibinfo {author}
  {\bibfnamefont {X.}~\bibnamefont {Feng}}, \bibinfo {author} {\bibfnamefont
  {H.}~\bibnamefont {Lian}}, \bibinfo {author} {\bibfnamefont {H.~Q.}\
  \bibnamefont {Liu}}, \bibinfo {author} {\bibfnamefont {A.~D.}\ \bibnamefont
  {Liu}}, \bibinfo {author} {\bibfnamefont {M.~K.}\ \bibnamefont {Han}},
  \bibinfo {author} {\bibfnamefont {J.~Q.}\ \bibnamefont {Dong}}, \bibinfo
  {author} {\bibfnamefont {H.~H.}\ \bibnamefont {Wang}}, \bibinfo {author}
  {\bibfnamefont {J.~W.}\ \bibnamefont {Liu}}, \bibinfo {author} {\bibfnamefont
  {Q.}~\bibnamefont {Zang}}, \bibinfo {author} {\bibfnamefont {S.~X.}\
  \bibnamefont {Wang}}, \bibinfo {author} {\bibfnamefont {T.~F.}\ \bibnamefont
  {Zhou}}, \bibinfo {author} {\bibfnamefont {Y.~H.}\ \bibnamefont {Huang}},
  \bibinfo {author} {\bibfnamefont {L.~Q.}\ \bibnamefont {Hu}}, \bibinfo
  {author} {\bibfnamefont {C.}~\bibnamefont {Zhou}}, \bibinfo {author}
  {\bibfnamefont {H.~X.}\ \bibnamefont {Qu}}, \bibinfo {author} {\bibfnamefont
  {Y.}~\bibnamefont {Chen}}, \bibinfo {author} {\bibfnamefont {S.~Y.}\
  \bibnamefont {Lin}}, \bibinfo {author} {\bibfnamefont {B.}~\bibnamefont
  {Zhang}}, \bibinfo {author} {\bibfnamefont {J.~P.}\ \bibnamefont {Qian}},
  \bibinfo {author} {\bibfnamefont {J.~S.}\ \bibnamefont {Hu}}, \bibinfo
  {author} {\bibfnamefont {G.~S.}\ \bibnamefont {Xu}}, \bibinfo {author}
  {\bibfnamefont {J.~L.}\ \bibnamefont {Chen}}, \bibinfo {author}
  {\bibfnamefont {K.}~\bibnamefont {Lu}}, \bibinfo {author} {\bibfnamefont
  {F.~K.}\ \bibnamefont {Liu}}, \bibinfo {author} {\bibfnamefont {Y.~T.}\
  \bibnamefont {Song}}, \bibinfo {author} {\bibfnamefont {J.~G.}\ \bibnamefont
  {Li}}, \ and\ \bibinfo {author} {\bibfnamefont {X.~Z.}\ \bibnamefont
  {Gong}},\ }\href@noop {} {\bibfield  {journal} {\bibinfo  {journal} {Phys.
  Rev. Lett.}\ }\textbf {\bibinfo {volume} {128}},\ \bibinfo {pages} {085003}
  (\bibinfo {year} {2022})}\BibitemShut {NoStop}%
\bibitem [{\citenamefont {Ye}\ \emph {et~al.}(2022)\citenamefont {Ye},
  \citenamefont {Luo}, \citenamefont {Xiao}, \citenamefont {Pan}, \citenamefont
  {Wang}, \citenamefont {Huang}, \citenamefont {Zang}, \citenamefont {Chen},
  \citenamefont {Jin}, \citenamefont {Wang}, \citenamefont {Xiao},\ and\
  \citenamefont {Wang}}]{YeNF22}%
  \BibitemOpen
  \bibfield  {author} {\bibinfo {author} {\bibfnamefont {L.}~\bibnamefont
  {Ye}}, \bibinfo {author} {\bibfnamefont {Z.}~\bibnamefont {Luo}}, \bibinfo
  {author} {\bibfnamefont {X.}~\bibnamefont {Xiao}}, \bibinfo {author}
  {\bibfnamefont {C.}~\bibnamefont {Pan}}, \bibinfo {author} {\bibfnamefont
  {Y.}~\bibnamefont {Wang}}, \bibinfo {author} {\bibfnamefont {Y.}~\bibnamefont
  {Huang}}, \bibinfo {author} {\bibfnamefont {Q.}~\bibnamefont {Zang}},
  \bibinfo {author} {\bibfnamefont {F.}~\bibnamefont {Chen}}, \bibinfo {author}
  {\bibfnamefont {Y.}~\bibnamefont {Jin}}, \bibinfo {author} {\bibfnamefont
  {S.}~\bibnamefont {Wang}}, \bibinfo {author} {\bibfnamefont {B.}~\bibnamefont
  {Xiao}}, \ and\ \bibinfo {author} {\bibfnamefont {S.}~\bibnamefont {Wang}},\
  }\href@noop {} {\bibfield  {journal} {\bibinfo  {journal} {Nucl. Fusion}\
  }\textbf {\bibinfo {volume} {62}},\ \bibinfo {pages} {124002} (\bibinfo
  {year} {2022})}\BibitemShut {NoStop}%
\bibitem [{\citenamefont {Han}\ \emph {et~al.}(2022)\citenamefont {Han},
  \citenamefont {Park}, \citenamefont {Sung}, \citenamefont {Kang},
  \citenamefont {Lee}, \citenamefont {Chung}, \citenamefont {Hahm},
  \citenamefont {Kim}, \citenamefont {Park}, \citenamefont {Bak}, \citenamefont
  {Cha}, \citenamefont {Choi}, \citenamefont {Choi}, \citenamefont {Gwak},
  \citenamefont {Hahn}, \citenamefont {Jang}, \citenamefont {Lee},
  \citenamefont {Kim}, \citenamefont {Kim}, \citenamefont {Kim}, \citenamefont
  {Ko}, \citenamefont {Ko}, \citenamefont {Lee}, \citenamefont {Lee},
  \citenamefont {Lee}, \citenamefont {Lee}, \citenamefont {Lee}, \citenamefont
  {Lee}, \citenamefont {Park}, \citenamefont {Seo}, \citenamefont {Yang},
  \citenamefont {Yoon},\ and\ \citenamefont {Na}}]{HanNature22}%
  \BibitemOpen
  \bibfield  {author} {\bibinfo {author} {\bibfnamefont {H.}~\bibnamefont
  {Han}}, \bibinfo {author} {\bibfnamefont {S.~J.}\ \bibnamefont {Park}},
  \bibinfo {author} {\bibfnamefont {C.}~\bibnamefont {Sung}}, \bibinfo {author}
  {\bibfnamefont {J.}~\bibnamefont {Kang}}, \bibinfo {author} {\bibfnamefont
  {Y.~H.}\ \bibnamefont {Lee}}, \bibinfo {author} {\bibfnamefont
  {J.}~\bibnamefont {Chung}}, \bibinfo {author} {\bibfnamefont {T.~S.}\
  \bibnamefont {Hahm}}, \bibinfo {author} {\bibfnamefont {B.}~\bibnamefont
  {Kim}}, \bibinfo {author} {\bibfnamefont {J.~K.}\ \bibnamefont {Park}},
  \bibinfo {author} {\bibfnamefont {J.~G.}\ \bibnamefont {Bak}}, \bibinfo
  {author} {\bibfnamefont {M.~S.}\ \bibnamefont {Cha}}, \bibinfo {author}
  {\bibfnamefont {G.~J.}\ \bibnamefont {Choi}}, \bibinfo {author}
  {\bibfnamefont {M.~J.}\ \bibnamefont {Choi}}, \bibinfo {author}
  {\bibfnamefont {J.}~\bibnamefont {Gwak}}, \bibinfo {author} {\bibfnamefont
  {S.~H.}\ \bibnamefont {Hahn}}, \bibinfo {author} {\bibfnamefont
  {J.}~\bibnamefont {Jang}}, \bibinfo {author} {\bibfnamefont {K.~C.}\
  \bibnamefont {Lee}}, \bibinfo {author} {\bibfnamefont {J.~H.}\ \bibnamefont
  {Kim}}, \bibinfo {author} {\bibfnamefont {S.~K.}\ \bibnamefont {Kim}},
  \bibinfo {author} {\bibfnamefont {W.~C.}\ \bibnamefont {Kim}}, \bibinfo
  {author} {\bibfnamefont {J.}~\bibnamefont {Ko}}, \bibinfo {author}
  {\bibfnamefont {W.~H.}\ \bibnamefont {Ko}}, \bibinfo {author} {\bibfnamefont
  {C.~Y.}\ \bibnamefont {Lee}}, \bibinfo {author} {\bibfnamefont {J.~H.}\
  \bibnamefont {Lee}}, \bibinfo {author} {\bibfnamefont {J.~H.}\ \bibnamefont
  {Lee}}, \bibinfo {author} {\bibfnamefont {J.~K.}\ \bibnamefont {Lee}},
  \bibinfo {author} {\bibfnamefont {J.~P.}\ \bibnamefont {Lee}}, \bibinfo
  {author} {\bibfnamefont {K.~D.}\ \bibnamefont {Lee}}, \bibinfo {author}
  {\bibfnamefont {Y.~S.}\ \bibnamefont {Park}}, \bibinfo {author}
  {\bibfnamefont {J.}~\bibnamefont {Seo}}, \bibinfo {author} {\bibfnamefont
  {S.~M.}\ \bibnamefont {Yang}}, \bibinfo {author} {\bibfnamefont {S.~W.}\
  \bibnamefont {Yoon}}, \ and\ \bibinfo {author} {\bibfnamefont {Y.~S.}\
  \bibnamefont {Na}},\ }\href@noop {} {\bibfield  {journal} {\bibinfo
  {journal} {Nature}\ }\textbf {\bibinfo {volume} {609}},\ \bibinfo {pages}
  {269} (\bibinfo {year} {2022})}\BibitemShut {NoStop}%
\bibitem [{\citenamefont {Connor}\ \emph {et~al.}(2004)\citenamefont {Connor},
  \citenamefont {Fukuda}, \citenamefont {Garbet}, \citenamefont {Gormezano},
  \citenamefont {Mukhovatov}, \citenamefont {Wakatani}, \citenamefont {the
  ITB~database group}, \citenamefont {the ITPA topical group~on transport},\
  and\ \citenamefont {internal~barrier physics}}]{ConnorNF04}%
  \BibitemOpen
  \bibfield  {author} {\bibinfo {author} {\bibfnamefont {J.~W.}\ \bibnamefont
  {Connor}}, \bibinfo {author} {\bibfnamefont {T.}~\bibnamefont {Fukuda}},
  \bibinfo {author} {\bibfnamefont {X.}~\bibnamefont {Garbet}}, \bibinfo
  {author} {\bibfnamefont {C.}~\bibnamefont {Gormezano}}, \bibinfo {author}
  {\bibfnamefont {V.}~\bibnamefont {Mukhovatov}}, \bibinfo {author}
  {\bibfnamefont {M.}~\bibnamefont {Wakatani}}, \bibinfo {author} {\bibnamefont
  {the ITB~database group}}, \bibinfo {author} {\bibnamefont {the ITPA topical
  group~on transport}}, \ and\ \bibinfo {author} {\bibnamefont
  {internal~barrier physics}},\ }\href@noop {} {\bibfield  {journal} {\bibinfo
  {journal} {Nucl. Fusion}\ }\textbf {\bibinfo {volume} {44}},\ \bibinfo
  {pages} {R1} (\bibinfo {year} {2004})}\BibitemShut {NoStop}%
\bibitem [{\citenamefont {Mantica}\ \emph {et~al.}(2008)\citenamefont
  {Mantica}, \citenamefont {Corrigan}, \citenamefont {Garbet}, \citenamefont
  {Imbeaux}, \citenamefont {Lonnroth}, \citenamefont {Parail}, \citenamefont
  {Tala}, \citenamefont {Taroni}, \citenamefont {M.},\ and\ \citenamefont
  {Weisen}}]{Mantica-JET-08}%
  \BibitemOpen
  \bibfield  {author} {\bibinfo {author} {\bibfnamefont {P.}~\bibnamefont
  {Mantica}}, \bibinfo {author} {\bibfnamefont {G.}~\bibnamefont {Corrigan}},
  \bibinfo {author} {\bibfnamefont {X.}~\bibnamefont {Garbet}}, \bibinfo
  {author} {\bibfnamefont {F.}~\bibnamefont {Imbeaux}}, \bibinfo {author}
  {\bibfnamefont {J.}~\bibnamefont {Lonnroth}}, \bibinfo {author}
  {\bibfnamefont {V.}~\bibnamefont {Parail}}, \bibinfo {author} {\bibfnamefont
  {T.}~\bibnamefont {Tala}}, \bibinfo {author} {\bibfnamefont {A.}~\bibnamefont
  {Taroni}}, \bibinfo {author} {\bibfnamefont {V.}~\bibnamefont {M.}}, \ and\
  \bibinfo {author} {\bibfnamefont {H.}~\bibnamefont {Weisen}},\ }\href@noop {}
  {\bibfield  {journal} {\bibinfo  {journal} {Fusion Sci. Tech.}\ }\textbf
  {\bibinfo {volume} {53}},\ \bibinfo {pages} {1152} (\bibinfo {year}
  {2008})}\BibitemShut {NoStop}%
\bibitem [{\citenamefont {Ida}\ and\ \citenamefont {Fujita}(2018)}]{IdaPPCF18}%
  \BibitemOpen
  \bibfield  {author} {\bibinfo {author} {\bibfnamefont {K.}~\bibnamefont
  {Ida}}\ and\ \bibinfo {author} {\bibfnamefont {F.}~\bibnamefont {Fujita}},\
  }\href@noop {} {\bibfield  {journal} {\bibinfo  {journal} {Plasma Phys.
  Control. Fusion}\ }\textbf {\bibinfo {volume} {60}},\ \bibinfo {pages}
  {033001} (\bibinfo {year} {2018})}\BibitemShut {NoStop}%
\bibitem [{\citenamefont {Biglary}\ \emph {et~al.}(1990)\citenamefont
  {Biglary}, \citenamefont {Diamond},\ and\ \citenamefont
  {Terry}}]{BiglaryPFB90}%
  \BibitemOpen
  \bibfield  {author} {\bibinfo {author} {\bibfnamefont {H.}~\bibnamefont
  {Biglary}}, \bibinfo {author} {\bibfnamefont {P.~H.}\ \bibnamefont
  {Diamond}}, \ and\ \bibinfo {author} {\bibfnamefont {P.~W.}\ \bibnamefont
  {Terry}},\ }\href@noop {} {\bibfield  {journal} {\bibinfo  {journal} {Phys.
  Fluids B}\ }\textbf {\bibinfo {volume} {2}},\ \bibinfo {pages} {1} (\bibinfo
  {year} {1990})}\BibitemShut {NoStop}%
\bibitem [{\citenamefont {Romanelli}\ and\ \citenamefont
  {Zonca}(1993)}]{RomanelliPoP93}%
  \BibitemOpen
  \bibfield  {author} {\bibinfo {author} {\bibfnamefont {F.}~\bibnamefont
  {Romanelli}}\ and\ \bibinfo {author} {\bibfnamefont {F.}~\bibnamefont
  {Zonca}},\ }\href@noop {} {\bibfield  {journal} {\bibinfo  {journal} {Phys.
  Plasmas}\ }\textbf {\bibinfo {volume} {5}},\ \bibinfo {pages} {4081}
  (\bibinfo {year} {1993})}\BibitemShut {NoStop}%
\bibitem [{\citenamefont {Kishimoto}\ \emph {et~al.}(1998)\citenamefont
  {Kishimoto}, \citenamefont {Kim}, \citenamefont {Horton}, \citenamefont
  {Tajima}, \citenamefont {LeBrun},\ and\ \citenamefont
  {Shirai}}]{KishimotoPPCF98}%
  \BibitemOpen
  \bibfield  {author} {\bibinfo {author} {\bibfnamefont {Y.}~\bibnamefont
  {Kishimoto}}, \bibinfo {author} {\bibfnamefont {J.~Y.}\ \bibnamefont {Kim}},
  \bibinfo {author} {\bibfnamefont {W.}~\bibnamefont {Horton}}, \bibinfo
  {author} {\bibfnamefont {T.}~\bibnamefont {Tajima}}, \bibinfo {author}
  {\bibfnamefont {M.~J.}\ \bibnamefont {LeBrun}}, \ and\ \bibinfo {author}
  {\bibfnamefont {H.}~\bibnamefont {Shirai}},\ }\href@noop {} {\bibfield
  {journal} {\bibinfo  {journal} {Plasma Phys. Control. Fusion}\ }\textbf
  {\bibinfo {volume} {40}},\ \bibinfo {pages} {A663} (\bibinfo {year}
  {1998})}\BibitemShut {NoStop}%
\bibitem [{\citenamefont {Rogister}(2000)}]{RogisterPoP00}%
  \BibitemOpen
  \bibfield  {author} {\bibinfo {author} {\bibfnamefont {A.}~\bibnamefont
  {Rogister}},\ }\href@noop {} {\bibfield  {journal} {\bibinfo  {journal}
  {Phys. Plasmas}\ }\textbf {\bibinfo {volume} {7}},\ \bibinfo {pages} {5070}
  (\bibinfo {year} {2000})}\BibitemShut {NoStop}%
\bibitem [{\citenamefont {Connor}\ and\ \citenamefont
  {Hastie}(2004)}]{ConnorPPCF04}%
  \BibitemOpen
  \bibfield  {author} {\bibinfo {author} {\bibfnamefont {J.~W.}\ \bibnamefont
  {Connor}}\ and\ \bibinfo {author} {\bibfnamefont {J.}~\bibnamefont
  {Hastie}},\ }\href@noop {} {\bibfield  {journal} {\bibinfo  {journal} {Plasma
  Phys. Control. Fusion}\ }\textbf {\bibinfo {volume} {46}},\ \bibinfo {pages}
  {1501} (\bibinfo {year} {2004})}\BibitemShut {NoStop}%
\bibitem [{\citenamefont {Sakamoto}\ \emph {et~al.}(2001)\citenamefont
  {Sakamoto}, \citenamefont {Kamada}, \citenamefont {Ide}, \citenamefont
  {Fujita}, \citenamefont {Shirai}, \citenamefont {Takizuka}, \citenamefont
  {Koide}, \citenamefont {Fukuda}, \citenamefont {Oikawa}, \citenamefont
  {Suzuki}, \citenamefont {Shinohara}, \citenamefont {Yoshino},\ and\
  \citenamefont {Team}}]{Sakamoto-JT60U-NF01}%
  \BibitemOpen
  \bibfield  {author} {\bibinfo {author} {\bibfnamefont {Y.}~\bibnamefont
  {Sakamoto}}, \bibinfo {author} {\bibfnamefont {Y.}~\bibnamefont {Kamada}},
  \bibinfo {author} {\bibfnamefont {S.}~\bibnamefont {Ide}}, \bibinfo {author}
  {\bibfnamefont {T.}~\bibnamefont {Fujita}}, \bibinfo {author} {\bibfnamefont
  {H.}~\bibnamefont {Shirai}}, \bibinfo {author} {\bibfnamefont
  {T.}~\bibnamefont {Takizuka}}, \bibinfo {author} {\bibfnamefont
  {Y.}~\bibnamefont {Koide}}, \bibinfo {author} {\bibfnamefont
  {T.}~\bibnamefont {Fukuda}}, \bibinfo {author} {\bibfnamefont
  {T.}~\bibnamefont {Oikawa}}, \bibinfo {author} {\bibfnamefont
  {T.}~\bibnamefont {Suzuki}}, \bibinfo {author} {\bibfnamefont
  {K.}~\bibnamefont {Shinohara}}, \bibinfo {author} {\bibfnamefont
  {R.}~\bibnamefont {Yoshino}}, \ and\ \bibinfo {author} {\bibfnamefont {J.-.}\
  \bibnamefont {Team}},\ }\href@noop {} {\bibfield  {journal} {\bibinfo
  {journal} {Nucl. Fusion}\ }\textbf {\bibinfo {volume} {41}},\ \bibinfo
  {pages} {865} (\bibinfo {year} {2001})}\BibitemShut {NoStop}%
\bibitem [{\citenamefont {Lin}\ \emph {et~al.}(1998)\citenamefont {Lin},
  \citenamefont {Hahm}, \citenamefont {Lee}, \citenamefont {Tang},\ and\
  \citenamefont {White}}]{LinScience98}%
  \BibitemOpen
  \bibfield  {author} {\bibinfo {author} {\bibfnamefont {Z.}~\bibnamefont
  {Lin}}, \bibinfo {author} {\bibfnamefont {T.~S.}\ \bibnamefont {Hahm}},
  \bibinfo {author} {\bibfnamefont {W.~W.}\ \bibnamefont {Lee}}, \bibinfo
  {author} {\bibfnamefont {W.~M.}\ \bibnamefont {Tang}}, \ and\ \bibinfo
  {author} {\bibfnamefont {R.~B.}\ \bibnamefont {White}},\ }\href@noop {}
  {\bibfield  {journal} {\bibinfo  {journal} {Science}\ }\textbf {\bibinfo
  {volume} {281}},\ \bibinfo {pages} {1835} (\bibinfo {year}
  {1998})}\BibitemShut {NoStop}%
\bibitem [{\citenamefont {McMillan}\ \emph {et~al.}(2009)\citenamefont
  {McMillan}, \citenamefont {Jolliet}, \citenamefont {Tran}, \citenamefont
  {Villard}, \citenamefont {Bottino},\ and\ \citenamefont
  {Angelino}}]{McMillanPoP09}%
  \BibitemOpen
  \bibfield  {author} {\bibinfo {author} {\bibfnamefont {B.~F.}\ \bibnamefont
  {McMillan}}, \bibinfo {author} {\bibfnamefont {S.}~\bibnamefont {Jolliet}},
  \bibinfo {author} {\bibfnamefont {T.~M.}\ \bibnamefont {Tran}}, \bibinfo
  {author} {\bibfnamefont {L.}~\bibnamefont {Villard}}, \bibinfo {author}
  {\bibfnamefont {A.}~\bibnamefont {Bottino}}, \ and\ \bibinfo {author}
  {\bibfnamefont {P.}~\bibnamefont {Angelino}},\ }\href@noop {} {\bibfield
  {journal} {\bibinfo  {journal} {Phys. Plasmas}\ }\textbf {\bibinfo {volume}
  {16}},\ \bibinfo {pages} {022310} (\bibinfo {year} {2009})}\BibitemShut
  {NoStop}%
\bibitem [{\citenamefont {Strugarek}\ \emph {et~al.}(2013)\citenamefont
  {Strugarek}, \citenamefont {Sarazin}, \citenamefont {Zarzoso}, \citenamefont
  {Abiteboul}, \citenamefont {Brun}, \citenamefont {Cartier-Michaud},
  \citenamefont {Dif-Pradalier}, \citenamefont {Garbet}, \citenamefont
  {Ghendrih}, \citenamefont {Grandgirard}, \citenamefont {Laru},\ and\
  \citenamefont {Thomine}}]{StrugarekPRL13}%
  \BibitemOpen
  \bibfield  {author} {\bibinfo {author} {\bibfnamefont {A.}~\bibnamefont
  {Strugarek}}, \bibinfo {author} {\bibfnamefont {Y.}~\bibnamefont {Sarazin}},
  \bibinfo {author} {\bibfnamefont {D.}~\bibnamefont {Zarzoso}}, \bibinfo
  {author} {\bibfnamefont {J.}~\bibnamefont {Abiteboul}}, \bibinfo {author}
  {\bibfnamefont {A.~S.}\ \bibnamefont {Brun}}, \bibinfo {author}
  {\bibfnamefont {T.}~\bibnamefont {Cartier-Michaud}}, \bibinfo {author}
  {\bibfnamefont {G.}~\bibnamefont {Dif-Pradalier}}, \bibinfo {author}
  {\bibfnamefont {X.}~\bibnamefont {Garbet}}, \bibinfo {author} {\bibfnamefont
  {P.}~\bibnamefont {Ghendrih}}, \bibinfo {author} {\bibfnamefont
  {V.}~\bibnamefont {Grandgirard}}, \bibinfo {author} {\bibfnamefont
  {G.}~\bibnamefont {Laru}}, \ and\ \bibinfo {author} {\bibfnamefont
  {O.}~\bibnamefont {Thomine}},\ }\href@noop {} {\bibfield  {journal} {\bibinfo
   {journal} {Phys. Ref. Lett.}\ }\textbf {\bibinfo {volume} {111}},\ \bibinfo
  {pages} {145001} (\bibinfo {year} {2013})}\BibitemShut {NoStop}%
\bibitem [{\citenamefont {Chen}\ \emph {et~al.}(2000)\citenamefont {Chen},
  \citenamefont {Lin},\ and\ \citenamefont {White}}]{ChenPoP00}%
  \BibitemOpen
  \bibfield  {author} {\bibinfo {author} {\bibfnamefont {L.}~\bibnamefont
  {Chen}}, \bibinfo {author} {\bibfnamefont {Z.}~\bibnamefont {Lin}}, \ and\
  \bibinfo {author} {\bibfnamefont {R.}~\bibnamefont {White}},\ }\href@noop {}
  {\bibfield  {journal} {\bibinfo  {journal} {Phys. Plasmas}\ }\textbf
  {\bibinfo {volume} {7}},\ \bibinfo {pages} {3129} (\bibinfo {year}
  {2000})}\BibitemShut {NoStop}%
\bibitem [{\citenamefont {Diamond}\ \emph {et~al.}(1994)\citenamefont
  {Diamond}, \citenamefont {Liang}, \citenamefont {Carreras},\ and\
  \citenamefont {Terry}}]{DiamondPRL94}%
  \BibitemOpen
  \bibfield  {author} {\bibinfo {author} {\bibfnamefont {P.~H.}\ \bibnamefont
  {Diamond}}, \bibinfo {author} {\bibfnamefont {Y.~L.}\ \bibnamefont {Liang}},
  \bibinfo {author} {\bibfnamefont {B.~A.}\ \bibnamefont {Carreras}}, \ and\
  \bibinfo {author} {\bibfnamefont {P.~W.}\ \bibnamefont {Terry}},\ }\href@noop
  {} {\bibfield  {journal} {\bibinfo  {journal} {Phys. Rev. Lett.}\ }\textbf
  {\bibinfo {volume} {72}},\ \bibinfo {pages} {2565} (\bibinfo {year}
  {1994})}\BibitemShut {NoStop}%
\bibitem [{\citenamefont {Wang}(2017)}]{WangPoP17}%
  \BibitemOpen
  \bibfield  {author} {\bibinfo {author} {\bibfnamefont {S.}~\bibnamefont
  {Wang}},\ }\href@noop {} {\bibfield  {journal} {\bibinfo  {journal} {Phys.
  Plasmas}\ }\textbf {\bibinfo {volume} {24}},\ \bibinfo {pages} {102508}
  (\bibinfo {year} {2017})}\BibitemShut {NoStop}%
\bibitem [{\citenamefont {Dai}\ \emph {et~al.}(2019)\citenamefont {Dai},
  \citenamefont {Xu}, \citenamefont {Ye}, \citenamefont {Xiao},\ and\
  \citenamefont {Wang}}]{DaiCPC19}%
  \BibitemOpen
  \bibfield  {author} {\bibinfo {author} {\bibfnamefont {Z.}~\bibnamefont
  {Dai}}, \bibinfo {author} {\bibfnamefont {Y.}~\bibnamefont {Xu}}, \bibinfo
  {author} {\bibfnamefont {L.}~\bibnamefont {Ye}}, \bibinfo {author}
  {\bibfnamefont {X.}~\bibnamefont {Xiao}}, \ and\ \bibinfo {author}
  {\bibfnamefont {S.}~\bibnamefont {Wang}},\ }\href@noop {} {\bibfield
  {journal} {\bibinfo  {journal} {Comput. Phys. Commun.}\ }\textbf {\bibinfo
  {volume} {242}},\ \bibinfo {pages} {72} (\bibinfo {year} {2019})}\BibitemShut
  {NoStop}%
\bibitem [{\citenamefont {Idomura}\ \emph {et~al.}(2008)\citenamefont
  {Idomura}, \citenamefont {Ida}, \citenamefont {Kano}, \citenamefont {Aiba},\
  and\ \citenamefont {Tokuda}}]{Idomura-GT5D-CPC08}%
  \BibitemOpen
  \bibfield  {author} {\bibinfo {author} {\bibfnamefont {Y.}~\bibnamefont
  {Idomura}}, \bibinfo {author} {\bibfnamefont {M.}~\bibnamefont {Ida}},
  \bibinfo {author} {\bibfnamefont {T.}~\bibnamefont {Kano}}, \bibinfo {author}
  {\bibfnamefont {N.}~\bibnamefont {Aiba}}, \ and\ \bibinfo {author}
  {\bibfnamefont {S.}~\bibnamefont {Tokuda}},\ }\href@noop {} {\bibfield
  {journal} {\bibinfo  {journal} {Comput. Phys. Commun.}\ }\textbf {\bibinfo
  {volume} {179}},\ \bibinfo {pages} {391} (\bibinfo {year}
  {2008})}\BibitemShut {NoStop}%
\bibitem [{\citenamefont {Jolliet}\ \emph {et~al.}(2007)\citenamefont
  {Jolliet}, \citenamefont {Bottino}, \citenamefont {Angelino}, \citenamefont
  {Hatzky}, \citenamefont {Tran}, \citenamefont {McMillan}, \citenamefont
  {Sauter}, \citenamefont {Appert}, \citenamefont {Idomura},\ and\
  \citenamefont {Villard}}]{Jolliet-ORB5-CPC07}%
  \BibitemOpen
  \bibfield  {author} {\bibinfo {author} {\bibfnamefont {S.}~\bibnamefont
  {Jolliet}}, \bibinfo {author} {\bibfnamefont {A.}~\bibnamefont {Bottino}},
  \bibinfo {author} {\bibfnamefont {P.}~\bibnamefont {Angelino}}, \bibinfo
  {author} {\bibfnamefont {R.}~\bibnamefont {Hatzky}}, \bibinfo {author}
  {\bibfnamefont {T.~M.}\ \bibnamefont {Tran}}, \bibinfo {author}
  {\bibfnamefont {B.~F.}\ \bibnamefont {McMillan}}, \bibinfo {author}
  {\bibfnamefont {O.}~\bibnamefont {Sauter}}, \bibinfo {author} {\bibfnamefont
  {K.}~\bibnamefont {Appert}}, \bibinfo {author} {\bibfnamefont
  {Y.}~\bibnamefont {Idomura}}, \ and\ \bibinfo {author} {\bibfnamefont
  {L.}~\bibnamefont {Villard}},\ }\href@noop {} {\bibfield  {journal} {\bibinfo
   {journal} {Comput. Phys. Commun.}\ }\textbf {\bibinfo {volume} {177}},\
  \bibinfo {pages} {409} (\bibinfo {year} {2007})}\BibitemShut {NoStop}%
\bibitem [{\citenamefont {Imadera}\ and\ \citenamefont
  {Kishimoto}(2023)}]{ImaderaPPCF23}%
  \BibitemOpen
  \bibfield  {author} {\bibinfo {author} {\bibfnamefont {K.}~\bibnamefont
  {Imadera}}\ and\ \bibinfo {author} {\bibfnamefont {Y.}~\bibnamefont
  {Kishimoto}},\ }\href@noop {} {\bibfield  {journal} {\bibinfo  {journal}
  {Plasma Phys. Control. Fusion}\ }\textbf {\bibinfo {volume} {65}},\ \bibinfo
  {pages} {024003} (\bibinfo {year} {2023})}\BibitemShut {NoStop}%
\bibitem [{\citenamefont {Brizard}\ and\ \citenamefont
  {Hahm}(2007)}]{BrizardRMP07}%
  \BibitemOpen
  \bibfield  {author} {\bibinfo {author} {\bibfnamefont {A.~J.}\ \bibnamefont
  {Brizard}}\ and\ \bibinfo {author} {\bibfnamefont {T.~S.}\ \bibnamefont
  {Hahm}},\ }\href@noop {} {\bibfield  {journal} {\bibinfo  {journal} {Rev.
  Mod. Phys.}\ }\textbf {\bibinfo {volume} {79}},\ \bibinfo {pages} {421}
  (\bibinfo {year} {2007})}\BibitemShut {NoStop}%
\bibitem [{\citenamefont {Wesson}(1997)}]{WessonBook97}%
  \BibitemOpen
  \bibfield  {author} {\bibinfo {author} {\bibfnamefont {J.}~\bibnamefont
  {Wesson}},\ }\href@noop {} {\emph {\bibinfo {title} {Tokamaks}}},\ \bibinfo
  {edition} {2nd}\ ed.\ (\bibinfo  {publisher} {Clarendon Press, Oxford},\
  \bibinfo {year} {1997})\ Chap.~\bibinfo {chapter} {4}\BibitemShut {NoStop}%
\bibitem [{\citenamefont {Wang}(2001)}]{WangPRE01}%
  \BibitemOpen
  \bibfield  {author} {\bibinfo {author} {\bibfnamefont {S.}~\bibnamefont
  {Wang}},\ }\href@noop {} {\bibfield  {journal} {\bibinfo  {journal} {Phys.
  Rev. E}\ }\textbf {\bibinfo {volume} {64}},\ \bibinfo {pages} {056404}
  (\bibinfo {year} {2001})}\BibitemShut {NoStop}%
\bibitem [{\citenamefont {Idomura}\ \emph {et~al.}(2003)\citenamefont
  {Idomura}, \citenamefont {Tokuda},\ and\ \citenamefont
  {Kishimoto}}]{IdomuraNF03}%
  \BibitemOpen
  \bibfield  {author} {\bibinfo {author} {\bibfnamefont {Y.}~\bibnamefont
  {Idomura}}, \bibinfo {author} {\bibfnamefont {S.}~\bibnamefont {Tokuda}}, \
  and\ \bibinfo {author} {\bibfnamefont {Y.}~\bibnamefont {Kishimoto}},\
  }\href@noop {} {\bibfield  {journal} {\bibinfo  {journal} {Nucl. Fusion}\
  }\textbf {\bibinfo {volume} {43}},\ \bibinfo {pages} {234} (\bibinfo {year}
  {2003})}\BibitemShut {NoStop}%
\bibitem [{\citenamefont {Ye}\ \emph {et~al.}(2016)\citenamefont {Ye},
  \citenamefont {Xu}, \citenamefont {Xiao}, \citenamefont {Dai},\ and\
  \citenamefont {Wang}}]{YeJCP16}%
  \BibitemOpen
  \bibfield  {author} {\bibinfo {author} {\bibfnamefont {L.}~\bibnamefont
  {Ye}}, \bibinfo {author} {\bibfnamefont {Y.}~\bibnamefont {Xu}}, \bibinfo
  {author} {\bibfnamefont {X.}~\bibnamefont {Xiao}}, \bibinfo {author}
  {\bibfnamefont {Z.}~\bibnamefont {Dai}}, \ and\ \bibinfo {author}
  {\bibfnamefont {S.}~\bibnamefont {Wang}},\ }\href@noop {} {\bibfield
  {journal} {\bibinfo  {journal} {J. Comput. Phys.}\ }\textbf {\bibinfo
  {volume} {316}},\ \bibinfo {pages} {180} (\bibinfo {year}
  {2016})}\BibitemShut {NoStop}%
\bibitem [{\citenamefont {Xu}\ \emph {et~al.}(2017)\citenamefont {Xu},
  \citenamefont {Ye}, \citenamefont {Dai}, , \citenamefont {Xiao},\ and\
  \citenamefont {Wang}}]{XuPoP17}%
  \BibitemOpen
  \bibfield  {author} {\bibinfo {author} {\bibfnamefont {Y.}~\bibnamefont
  {Xu}}, \bibinfo {author} {\bibfnamefont {L.}~\bibnamefont {Ye}}, \bibinfo
  {author} {\bibfnamefont {Z.}~\bibnamefont {Dai}}, , \bibinfo {author}
  {\bibfnamefont {Z.}~\bibnamefont {Xiao}}, \ and\ \bibinfo {author}
  {\bibfnamefont {S.}~\bibnamefont {Wang}},\ }\href@noop {} {\bibfield
  {journal} {\bibinfo  {journal} {Phys. Plasmas}\ }\textbf {\bibinfo {volume}
  {24}},\ \bibinfo {pages} {082515} (\bibinfo {year} {2017})}\BibitemShut
  {NoStop}%
\bibitem [{\citenamefont {Wang}(2012)}]{WangPoP12}%
  \BibitemOpen
  \bibfield  {author} {\bibinfo {author} {\bibfnamefont {S.}~\bibnamefont
  {Wang}},\ }\href@noop {} {\bibfield  {journal} {\bibinfo  {journal} {Phys.
  Plasmas}\ }\textbf {\bibinfo {volume} {19}},\ \bibinfo {pages} {062504}
  (\bibinfo {year} {2012})}\BibitemShut {NoStop}%
\bibitem [{\citenamefont {Wang}(2013)}]{WangPoP13}%
  \BibitemOpen
  \bibfield  {author} {\bibinfo {author} {\bibfnamefont {S.}~\bibnamefont
  {Wang}},\ }\href@noop {} {\bibfield  {journal} {\bibinfo  {journal} {Phys.
  Plasmas}\ }\textbf {\bibinfo {volume} {20}},\ \bibinfo {pages} {082312}
  (\bibinfo {year} {2013})}\BibitemShut {NoStop}%
\bibitem [{\citenamefont {Lee}(1987)}]{LeeJCP87}%
  \BibitemOpen
  \bibfield  {author} {\bibinfo {author} {\bibfnamefont {W.~W.}\ \bibnamefont
  {Lee}},\ }\href@noop {} {\bibfield  {journal} {\bibinfo  {journal} {J.
  Comput. Phys.}\ }\textbf {\bibinfo {volume} {72}},\ \bibinfo {pages} {243}
  (\bibinfo {year} {1987})}\BibitemShut {NoStop}%
\bibitem [{\citenamefont {Chen}\ and\ \citenamefont {Zonca}(2007)}]{ChenNF07}%
  \BibitemOpen
  \bibfield  {author} {\bibinfo {author} {\bibfnamefont {L.}~\bibnamefont
  {Chen}}\ and\ \bibinfo {author} {\bibfnamefont {F.}~\bibnamefont {Zonca}},\
  }\href@noop {} {\bibfield  {journal} {\bibinfo  {journal} {Nucl. Fusion}\
  }\textbf {\bibinfo {volume} {47}},\ \bibinfo {pages} {886} (\bibinfo {year}
  {2007})}\BibitemShut {NoStop}%
\bibitem [{\citenamefont {Bottino}\ \emph {et~al.}(2022)\citenamefont
  {Bottino}, \citenamefont {Falessi}, \citenamefont {Hayward-Schneider},
  \citenamefont {Biancalani}, \citenamefont {Briguglio}, \citenamefont
  {Hatzky}, \citenamefont {Lauber}, \citenamefont {Mishchenko}, \citenamefont
  {Poli}, \citenamefont {Rettino}, \citenamefont {Vannini}, \citenamefont
  {Wang},\ and\ \citenamefont {Zonca}}]{BottinoJoP22}%
  \BibitemOpen
  \bibfield  {author} {\bibinfo {author} {\bibfnamefont {A.}~\bibnamefont
  {Bottino}}, \bibinfo {author} {\bibfnamefont {M.~V.}\ \bibnamefont
  {Falessi}}, \bibinfo {author} {\bibfnamefont {T.}~\bibnamefont
  {Hayward-Schneider}}, \bibinfo {author} {\bibfnamefont {A.}~\bibnamefont
  {Biancalani}}, \bibinfo {author} {\bibfnamefont {S.}~\bibnamefont
  {Briguglio}}, \bibinfo {author} {\bibfnamefont {R.}~\bibnamefont {Hatzky}},
  \bibinfo {author} {\bibfnamefont {P.}~\bibnamefont {Lauber}}, \bibinfo
  {author} {\bibfnamefont {A.}~\bibnamefont {Mishchenko}}, \bibinfo {author}
  {\bibfnamefont {E.}~\bibnamefont {Poli}}, \bibinfo {author} {\bibfnamefont
  {B.}~\bibnamefont {Rettino}}, \bibinfo {author} {\bibfnamefont
  {F.}~\bibnamefont {Vannini}}, \bibinfo {author} {\bibfnamefont
  {X.}~\bibnamefont {Wang}}, \ and\ \bibinfo {author} {\bibfnamefont
  {F.}~\bibnamefont {Zonca}},\ }\href@noop {} {\bibfield  {journal} {\bibinfo
  {journal} {J. Physics: Conference Series}\ }\textbf {\bibinfo {volume}
  {2397}},\ \bibinfo {pages} {012019} (\bibinfo {year} {2022})}\BibitemShut
  {NoStop}%
\bibitem [{\citenamefont {Garbet}\ \emph {et~al.}(2001)\citenamefont {Garbet},
  \citenamefont {Bourdelle}, \citenamefont {Hoang}, \citenamefont {Maget},
  \citenamefont {Benkadda}, \citenamefont {Beyer}, \citenamefont {Figarella},
  \citenamefont {Voitsekovitch}, \citenamefont {Agullo},\ and\ \citenamefont
  {Bian}}]{GarbetPoP01}%
  \BibitemOpen
  \bibfield  {author} {\bibinfo {author} {\bibfnamefont {X.}~\bibnamefont
  {Garbet}}, \bibinfo {author} {\bibfnamefont {C.}~\bibnamefont {Bourdelle}},
  \bibinfo {author} {\bibfnamefont {G.~T.}\ \bibnamefont {Hoang}}, \bibinfo
  {author} {\bibfnamefont {P.}~\bibnamefont {Maget}}, \bibinfo {author}
  {\bibfnamefont {S.}~\bibnamefont {Benkadda}}, \bibinfo {author}
  {\bibfnamefont {P.}~\bibnamefont {Beyer}}, \bibinfo {author} {\bibfnamefont
  {C.}~\bibnamefont {Figarella}}, \bibinfo {author} {\bibfnamefont
  {I.}~\bibnamefont {Voitsekovitch}}, \bibinfo {author} {\bibfnamefont
  {O.}~\bibnamefont {Agullo}}, \ and\ \bibinfo {author} {\bibfnamefont
  {N.}~\bibnamefont {Bian}},\ }\href@noop {} {\bibfield  {journal} {\bibinfo
  {journal} {Phys. Plasmas}\ }\textbf {\bibinfo {volume} {8}},\ \bibinfo
  {pages} {2793} (\bibinfo {year} {2001})}\BibitemShut {NoStop}%
\bibitem [{\citenamefont {Wagner}(2007)}]{WagnerPPCF07}%
  \BibitemOpen
  \bibfield  {author} {\bibinfo {author} {\bibfnamefont {F.}~\bibnamefont
  {Wagner}},\ }\href@noop {} {\bibfield  {journal} {\bibinfo  {journal} {Plasma
  Phys. Control. Fusion}\ }\textbf {\bibinfo {volume} {49}},\ \bibinfo {pages}
  {B1} (\bibinfo {year} {2007})}\BibitemShut {NoStop}%
\bibitem [{\citenamefont {Barnes}\ \emph {et~al.}(2010)\citenamefont {Barnes},
  \citenamefont {Abel}, \citenamefont {Dorland}, \citenamefont {Gorler},
  \citenamefont {Hammett},\ and\ \citenamefont {Jenko}}]{BarnesPoP10}%
  \BibitemOpen
  \bibfield  {author} {\bibinfo {author} {\bibfnamefont {M.}~\bibnamefont
  {Barnes}}, \bibinfo {author} {\bibfnamefont {I.~G.}\ \bibnamefont {Abel}},
  \bibinfo {author} {\bibfnamefont {W.}~\bibnamefont {Dorland}}, \bibinfo
  {author} {\bibfnamefont {T.}~\bibnamefont {Gorler}}, \bibinfo {author}
  {\bibfnamefont {G.~W.}\ \bibnamefont {Hammett}}, \ and\ \bibinfo {author}
  {\bibfnamefont {F.}~\bibnamefont {Jenko}},\ }\href@noop {} {\bibfield
  {journal} {\bibinfo  {journal} {Phys. Plasmas}\ }\textbf {\bibinfo {volume}
  {17}},\ \bibinfo {pages} {056109} (\bibinfo {year} {2010})}\BibitemShut
  {NoStop}%
\end{thebibliography}

%

\end{document}